\documentclass[%
 reprint,
 amsmath,amssymb,
 aps,
]{revtex4-2}
\usepackage{graphicx}
\usepackage{dcolumn}
\usepackage{bm}
\usepackage{xcolor}

\usepackage{caption} 
\usepackage{subcaption} 

\usepackage{comment}

\begin{document}

\preprint{APS/123-QED}

\title{Versatile, high brightness, cryogenic photoinjector electron source}

\author{River R. Robles}
 \email{riverr@stanford.edu}
 \altaffiliation[Now at ]{Department of Applied Physics, Stanford University, Stanford, CA 94305. }
\author{Obed Camacho}
\author{Atsushi Fukasawa}
\author{Nathan Majernik}
\author{James B. Rosenzweig}%
\affiliation{%
 Department of Physics and Astronomy, University of California, Los Angeles, 405 Hilgard Ave., Los Angeles, CA 90095, USA
}%

\date{\today}

\begin{abstract}
Since the introduction of the radio-frequency (rf) photoinjector electron source over thirty years ago, peak performance demands have dictated the use of high accelerating electric fields. With recent strong advances in obtainable field values, attendant increases in beam brightness are expected to be dramatic. In this article, we examine the implementation of very high gradient acceleration in a high frequency, cryogenic rf photoinjector. We discuss in detail the effects of introducing, through an optimized rf cavity shape, rich spatial harmonic content in the accelerating modes in this device. Higher spatial harmonics give useful, enhanced linear focusing effects, as well as potentially deleterious nonlinear transverse forces. They also serve to strongly increase the ratio of average accelerating field to peak surface field, thus aiding in managing power and dark current-related challenges. We investigate two scenarios which are aimed at unique exploitation of the capabilities of this source. First, we investigate the obtaining of extremely high six-dimensional brightness for advanced free-electron laser applications.  We also examine the use of a magnetized photocathode in the device for producing unprecedented low asymmetric emittance, high-current electron beams that reach linear collider-compatible performance. As both of the scenarios demand an advanced, compact solenoid design, we describe a novel cryogenic solenoid system. With the high field rf and magnetostatic structures introduced, we analyze the collective beam dynamics in these systems through theory and multi-particle simulations, including a particular emphasis on granularity effects associated with microscopic Coulomb interactions. 

\end{abstract}

\maketitle

\section{Introduction}

The radio-frequency (rf) photoinjector is a class of electron source that has transformed numerous fields in beam-based science, providing relativistic electron beams with unprecedented short pulse length, high current, and low emittance. This is accomplished by laser-gated photo-emission from a cathode embedded in a very high field rf cavity, liberating picosecond or faster electron pulses through a prompt (as low as tens of femtosecond delay) photoelectric emission. It is after emission that the true innovation of the photoinjector begins, however, as if the beam was simply accelerated from the cathode with no additional optics it would experience strong correlated emittance growth due to current-dependent transverse space-charge forces. However, it was shown by Carlsten \cite{CARLSTEN1989313} that such correlated emittance growth could be reversed by focusing the beam soon after it emerges from the cathode, and one can in fact retrieve the emittance the beam was born with. This concept was developed further by Serafini and Rosenzweig \cite{serafini1997envelope}, where the process of undoing current-dependent correlations, deemed \textit{emittance compensation}, became a prescribed process with well-understood working points to seek out for optimal compensation. The importance of these developments is straightforward to appreciate -- they enabled the robust performance of the world's first x-ray free-electron laser (XFEL) \cite{emma2010first}. 

The first generation of photoinjectors used in large facilities provided beams which were of sufficient quality to enable the world's first XFEL.  Today, however, we find many innovative new applications for electron beams which demand ever-brighter sources. Unsurprisingly, these innovations in the use of high-brightness beams present a concomitant challenge, to strongly increase the beam brightness produced by the source. In regard to XFEL applications, two recent initiatives indicate a necessity for beams which far exceed the current state-of-the-art: the ultra-compact XFEL \cite{rosenzweig2020ultra} and the MaRIE XFEL \cite{carlsten2019high}. The first, pioneered by a UCLA-centered collaboration, is an ultra-compact XFEL (UC-XFEL), which promises lasing, initially at soft x-ray wavelengths, with a total footprint under 40 m in length. The fundamental feature allowing the substantially shorter total length is the reduction of the final beam energy, from the several GeV level down to 1 GeV. Doing so without dramatically reducing the FEL power requires maintaining the geometric emittance of the beam at the lower energy, thereby demanding a dramatic reduction of the normalized emittance at emission. Further insights into the demands on the electron beam brightness result from the analysis presented in \cite{rosenzweig2020ultra}, where it is shown that the key performance metric for source improvement is the six-dimensional brightness, which also includes the beam spectral density. 

The second emergent FEL application, the Matter Radiation Interactions in Extremes (MaRIE) XFEL, seeks to provide a photon source at extremely high photon energies exceeding 40 keV, or equivalently an x-ray wavelength below one-third of an Angstrom. Due to the stringent constraints imposed on beam emittance by the FEL instability, early design work towards such an FEL has struggled to identify a robust scenario, primarily due to degradation to beam brightness during transport to energies in excess of 10 GeV. Even designs which utilize 200 nm rad normalized emittance, at the current  state-of-the-art of frontier injectors, meet performance metrics but with uncomfortably little room for error. As such, this project would benefit greatly from an enhancement of the beam brightness at the source, as was demonstrated by the work reported in Ref. \cite{carlsten2019high}.

Like the x-ray free-electron laser, conceptual designs for a future electron-positron linear collider demand extremely high-brightness beams with even higher charge \cite{bambade2019international}. The beams required for a linear collider, however, go further than requiring innovative approaches to high-brightness beam development. In order to mitigate beam-beam radiation effects at the interaction point, known as \textit{beamsstrahlung}, one wishes for the beams to be transversely asymmetric, or ``flat", and in particular have asymmetric transverse emittances \cite{ChenTelnov}. Although there are several approaches to generating such beams, the only one capable of maintaining high brightness and high charge without excessive additions to footprint or cost is the photoinjector operated with a magnetized photocathode. This entails immersing the cathode in an axial magnetic field such that the electron beam is born with a non-zero canonical angular momentum which, as a conserved quantity, is converted to mechanical angular momentum downstream of the solenoid region. The presence of this mechanical angular momentum enables the splitting of the emittances via use of a skew quadrupole triplet \cite{KimRoundFlat}. In the case of the linear collider, we must maintain an ultra-high brightness beam while purposefully immersing the cathode in a magnetic field -- a situation which is traditionally avoided (\textit{e.g. }in FEL applications) due to its potentially damaging effects on the beam emittance.

Flat beams also find an application in dielectric laser acceleration (DLA) \cite{england2014dielectric}, particularly in structures with slab geometries, where, in order to generate extremely high accelerating gradients, the structures necessarily have gaps which are very small in one transverse dimension, and very large in the other \cite{Naranjo2012}\cite{Yousefi19}. The use of an un-magnetized photoinjector to produce these beams is hindered by the single-nm emittance requirements in the small dimension. However, if the emittance is split then the smaller transverse dimension can achieve single-nm level emittance without requiring the same level thermal emittance at the photocathode.

The proposal that the path towards ever-brighter beams lies in cryogenically-cooled normal conducting rf structures is at this point well-established in theory and simulation, as well as fundamental work on rf cavity performance \cite{CahillBreakdown}. Previous work on the subject \cite{rosenzweig2018ultra,rosenzweig2019next} has demonstrated this in the simplest terms: with a standard sinusoidal accelerating wave at the gun with 240 MV/m peak field, the beams that can be produced exceed the state-of-the-art by an order of magnitude in brightness. Beams with these performance metrics -- 55 nm rad normalized emittance, 20 A current, and sub-keV energy spread -- have already been used in simulation studies of the UC-XFEL and MaRIE XFEL demonstrating excellent performance. However, new developments in the burgeoning field of cryo-rf accelerating structures demand an updated treatment of such an injector. These new developments revolve largely around advances in the design of copper accelerating structures, both that of the accelerating cavities themselves as well as the systems used to couple rf power into them. The distributed coupling linac \cite{Tantawi2020} is unique among photoinjectors and accelerating structures alike in its square-wave-like field profile. The non-sinusoidal profile is indicative of the presence of higher harmonic content in the accelerating wave, the effects of which have been studied broadly in early studies of rf photoinjectors but have never been studied in the context of fields as high as we consider here, nor with beams as bright as those we present. 

The specific design requirements of the UC-XFEL have also recently become mature. Ref. \cite{rosenzweig2020ultra} presents a detailed summary of these requirements, which elucidated not just the  five-dimensional beam brightness $B_{\mathrm{5D}}=2I/\epsilon_n^2$ demanded, but extended the analysis to identify the needed six-dimensional brightness $B_{\mathrm{6D}}=2I/\epsilon_n^2\sigma_\gamma$. The recently published design study indicates the upper-bound on allowed energy spread at the injector, which necessitates a closer look at the physical processes that determine that energy spread. In particular, it demands the inclusion of the effects of microscopic space-charge effects associated with short-range Coulomb interactions such as intra-beam scattering (IBS). Capturing these effects in simulation is, however, extremely difficult as they demand a one-to-one treatment of the beam's collective effects at least between neighboring electrons. Such one-to-one simulations with a 100 pC beam associated with the applications analyzed here are computationally unwieldy, and demand a physics-based methodology for capturing the appropriate effects with a more realistic computational approach. 

In this article we will present a robust design for a cryogenically cooled rf photoinjector capable of meeting the ambitious brightness demands of next-generation electron beam applications for both symmetric and asymmetric emittance instruments. Although our primary emphasis will be on beam dynamics rather than injector engineering, this begins with a review of the physics which enables high-field acceleration in normal conducting, cryo-cooled copper, as well as the engineering designs of the gun, linac, and solenoid which allow us to realize these benefits experimentally. Once the engineering feasibility is reviewed, we present an in-depth analysis of the impact of higher harmonic field content in an injector -- in particular in the context of the present cell design and the high brightness of the beams, but also more generally with regard to modifications to linear beam dynamics and the strength of deleterious nonlinear forces. This work is based on early work on such effects and also builds on them, lending understanding to their role in the upcoming generation of high-brightness photoinjectors. After this, we will briefly review the theory of emittance compensation in standard non-magnetized injectors so as to develop the language and tools necessary to understand how compensation is modified when the cathode is immersed in a magnetic field. 

Once we have established the theoretical tools involved in the design of an injector in the present context, we will present multi-particle simulation studies of several relevant injector working points. The first, which we deem the ultra-high brightness working point, is a design extremely well-suited for driving a short-wavelength compact FEL, as it  yields a 19 A beam with 45 nm rad emittance and sub-keV energy spread. This predicted brightness performance notably exceeds that of even previously studied cryo-cooled guns. We then proceed to a study of this working point scaled down to a lower charge in order to facilitate one-to-one space-charge studies in a computationally feasible manner. This scaling, based on the observation that short-range Coulomb interactions scale with the beam charge density, allows us to estimate the energy spread from microscopic effects in the full 100 pC bunch using a beam with a charge, and therefore particle number, reduced by three orders of magnitude. Finally, we introduce an additional bucking solenoid element in the photoinjector design to permit cathode magnetization, and the associated production of asymmetric emittance beams. We will show that the four-dimensional beam brightness $\epsilon_{\mathrm{4D}}=\sqrt{\epsilon_x\epsilon_y}$ can be preserved down to nearly the level of the ultra-high brightness injector with a subsequent emittance-splitting by a ratio of 400. This yields a beam with 4 nm emittance in the smaller transverse plane at 100 pC bunch charge, which can be scaled to meet the demands of either a DLA or a linear collider. The physics involved in such a scaling procedure are also discussed. 

\section{Design of Photoinjector Components \label{sec:engineering}}

\subsection{Optimized rf structure design}

The use of cryogenic copper cavities to reach high electric fields is motivated first by material properties. At cryogenic temperatures, a number profound changes in material response are noted. First, the power dissipation due to surface currents is diminished strongly -- a factor of ~4-5 for relevant rf frequencies -- by entry into the anomalous skin effect regime. This causes the pulsed heating suffered by the cavity surface to be ameliorated. Second, the material properties - the coefficient of thermal expansion and, to a lesser extent, the thermal conductivity, change in beneficial ways that permit the deposited heat to produce less stress on the rf cavity surface. Finally, the yield strength is greatly increased at cryogenic temperatures, leading to a greater ability of the structure to withstand the impulse of the electric field and associated surface failure. Aspects of this microscopic model have been verified, with the role of the magnetic field (pulsed heating) \cite{Laurent2011} and the electric field \cite{Efieldbreakdown}, respectively, experimentally studied. The definitive study examining these effects jointly occurring in an accelerator-like cryogenic copper structure is found in Ref. \cite{CahillBreakdown}.  

\begin{figure}[h!]
    \centering
    \includegraphics[width=\columnwidth]{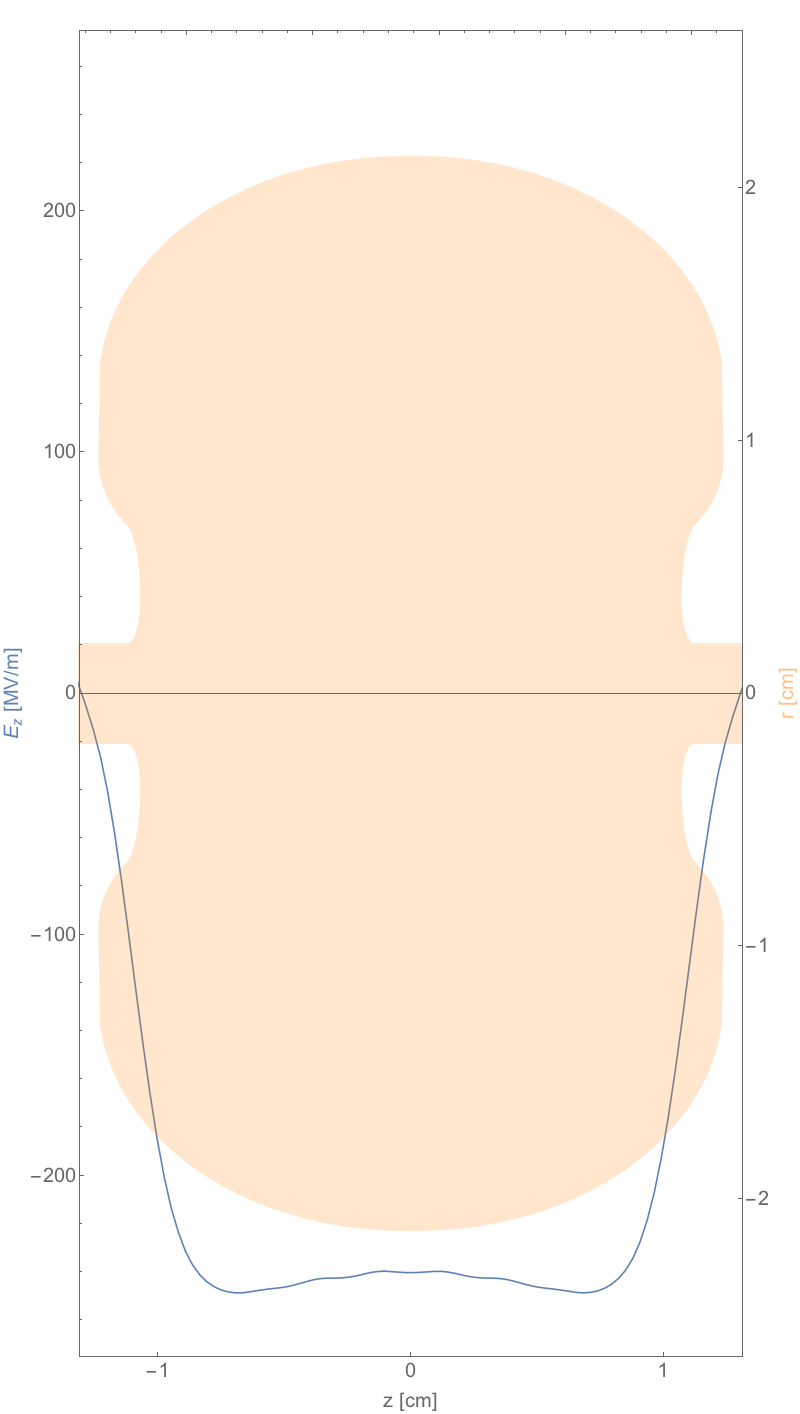}
    \caption{The axial field profile for the full cell design is plotted on top of a cross-section of the cell geometry with true-to-life aspect ratio.}
    \label{fig:fullcell_profile}
\end{figure}

Based on this last study, a proposal was made to employ this technique in very high field rf photoinjectors, as a way to increase the brightness. Indeed, the rf gun design we consider here is a more mature version of that first introduced in previous work \cite{rosenzweig2018ultra,rosenzweig2019next}, which examined a design with a field distribution very close to a pure standing wave $\pi$-mode with negligible spatial harmonic content. In the present work the field profile is modified from this pure standing wave as a result of a detailed optimization of the cell geometry, as is clear from Figure \ref{fig:fullcell_profile}. The optimized rf structure has a very high shunt impedance, owing both to the increased quality factor $Q$ and to a re-entrant geometry that is feasible in $\pi$-mode operation owing to a unique distributed coupling architecture \cite{Tantawi2020}. This structure geometry is motivated by the goals of minimizing the surface electric and magnetic field strengths. Taken together, achieving these goals results in a structure which can support peak on-axis electric fields in excess of 500 MV/m without suffering from excessive breakdown (owed to magnetic field-induced heating and electric field-driven stress). Further, we choose a peak design field for injector operation of 240 MV/m specifically to avoid dark current concerns, which are not a notable issue until the peak fields reach the threshold of 300 MV/m found in Ref. \cite{CahillRFloss}. The residual issues of dark current are planned to be managed in operations by limiting the rf pulse length and employing active sweeping methods. These are evaluated to be adequate solutions, based on the previous experimental results, for operation at the moderate (100 pC and above) beam charges examined in this paper. 

On the other side of the trade-off involved in this design, the field profile is populated by many higher spatial harmonics in addition to the primary resonant wave. This feature of the field profile was not accounted for in previous iterations of gun design, owing to the lack of robust, detailed rf design, and introduces several unique features to the emittance compensation process which we explore in this article. 

Although the present work's emphasis is in beam dynamics, we provide a more complete description of the rf characteristics in Table \ref{tab:rf}. The point of rf power requirements merits some additional clarification. Presently the two sections of the gun (full cell and 0.6 cell) are planned to be fed individually. Each section will be provided an initial 300 ns fill pulse followed by a subsequent 300 ns pulse for maintaining the fill during propagation of the beam. In Table \ref{tab:rf} the values quoted as ``A+B" should be understood as corresponding to the power contained in the first and second of these pulses in each gun section, respectively. Furthermore, the repetition rate of 100 Hz has been chosen primarily for the UC-XFEL context. At this rate, one dissipates 11 W at cryogenic temperatures, for an estimated cryocooler power of 500 W. This is thus easily feasible, and one can imagine pushing to higher repetition rates when only the gun operation is considered. One final point should be noted about the impact of rf loading. For the FEL case, there is no need to inject multiple bunches per rf fill, which is the context in which loading can become a problem. Multi-pulse operation is implicit in the linear collider case, however, in which case the small beam loading effects can be dealt with by adjusting the external rf feed. Additional details about these high gradient structures - their capabilities, manufacturing techniques, and recent results - can be found in references \cite{Tantawi2020, nasr2020experimental} which represent the most current state-of-the-art in published work about these cavities. Further details about the rf studies in progress and planned at UCLA will be provided in the conclusion. 

\begin{table}[h!]
    \centering
    \begin{tabular}{|c|c|c|}
        \hline
        Parameter & Unit & Value \\
        \hline
        Repetition Rate & Hz & 100\\
        rf Frequency & GHz & 5.712\\
        Operating Temperature & K & 27\\
        Input Power (FC) & MW & 10.7+3\\
        Input Power (0.6C) & MW & 4.8+1.6\\
        Dissipated Energy (FC) & J & 0.72 \\
        Dissipated Energy (0.6C) & J & 0.39\\
        Shunt Impedance & M$\Omega$/m & 121\\
        Pulse Length & ns & 300\\
        Quality Factor & & 14000\\
        \hline  
    \end{tabular}
    \caption{Several key rf design characteristics are reported. The two numbers quoted for each input power value correspond to the powers of the initial and secondary rf pulses, as explained in the text. }
    \label{tab:rf}
\end{table}

\subsection{Cryogenic solenoid}

\begin{figure}[htb]
     \centering
     \begin{subfigure}[b]{\columnwidth}
         \centering
         \includegraphics[width=\columnwidth]{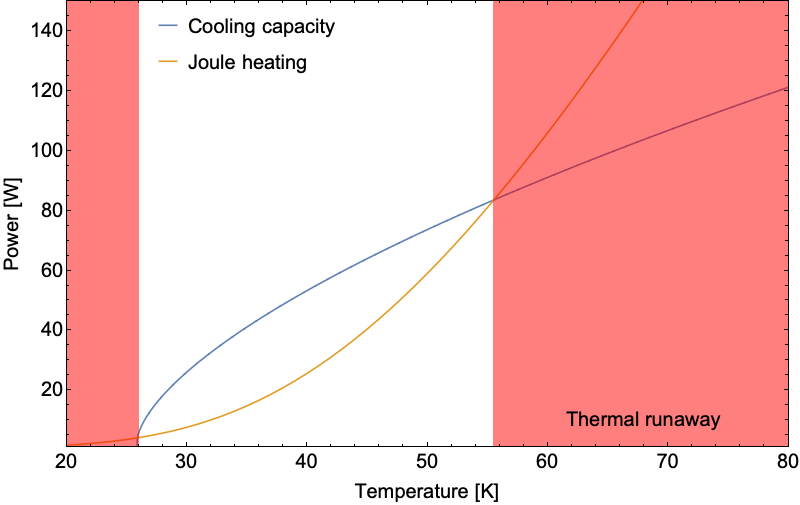}
         \caption{Power curves as a function of temperature for a representative, single stage Gifford-McMahon cold head and for the cryosolenoid, at a uniform temperature. Additional heat loads will reduce the effective cooling capacity.}
         \label{fig:cryosolenoid-powerCurves}
     \end{subfigure}
     \hfill
     \begin{subfigure}[b]{\columnwidth}
         \centering
         \includegraphics[width=\columnwidth]{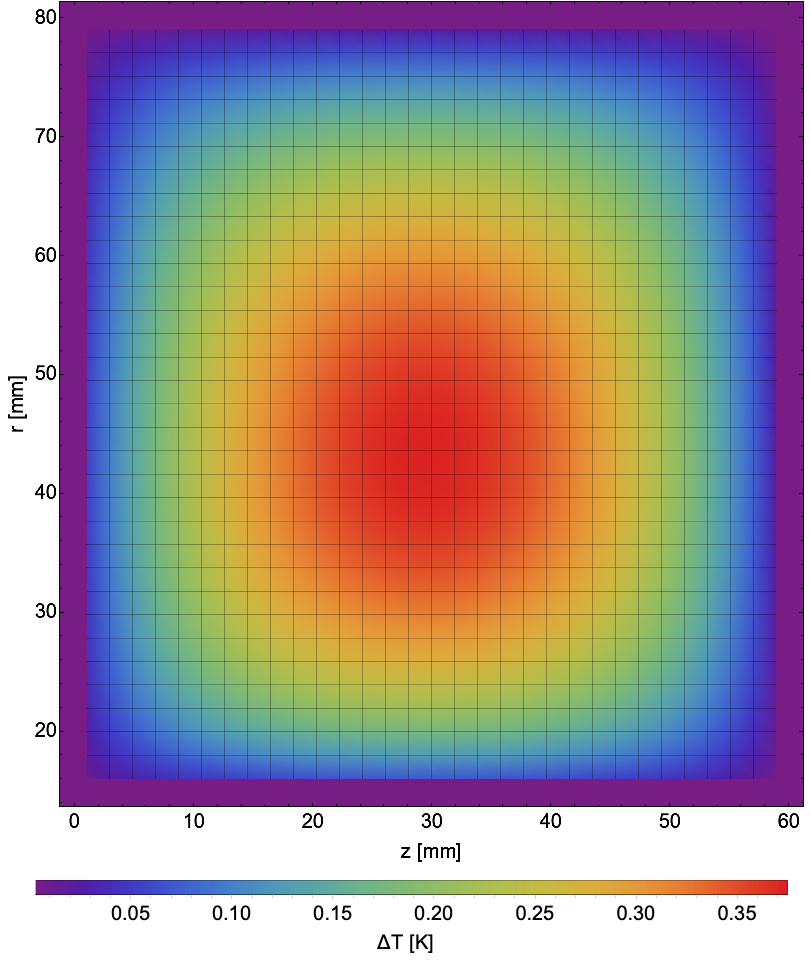}
         \caption{Equilibrium thermal distribution of winding cross section with 27 K sarcophagus assumed.}
         \label{fig:cryosolenoid-thermal}
     \end{subfigure}
     
    \caption{Plots of cryosolenoid thermal performance.}
    \label{fig:cryosolenoid-combined}
\end{figure}

\begin{figure*}[htb]
    \centering
    \includegraphics[width=0.75\textwidth]{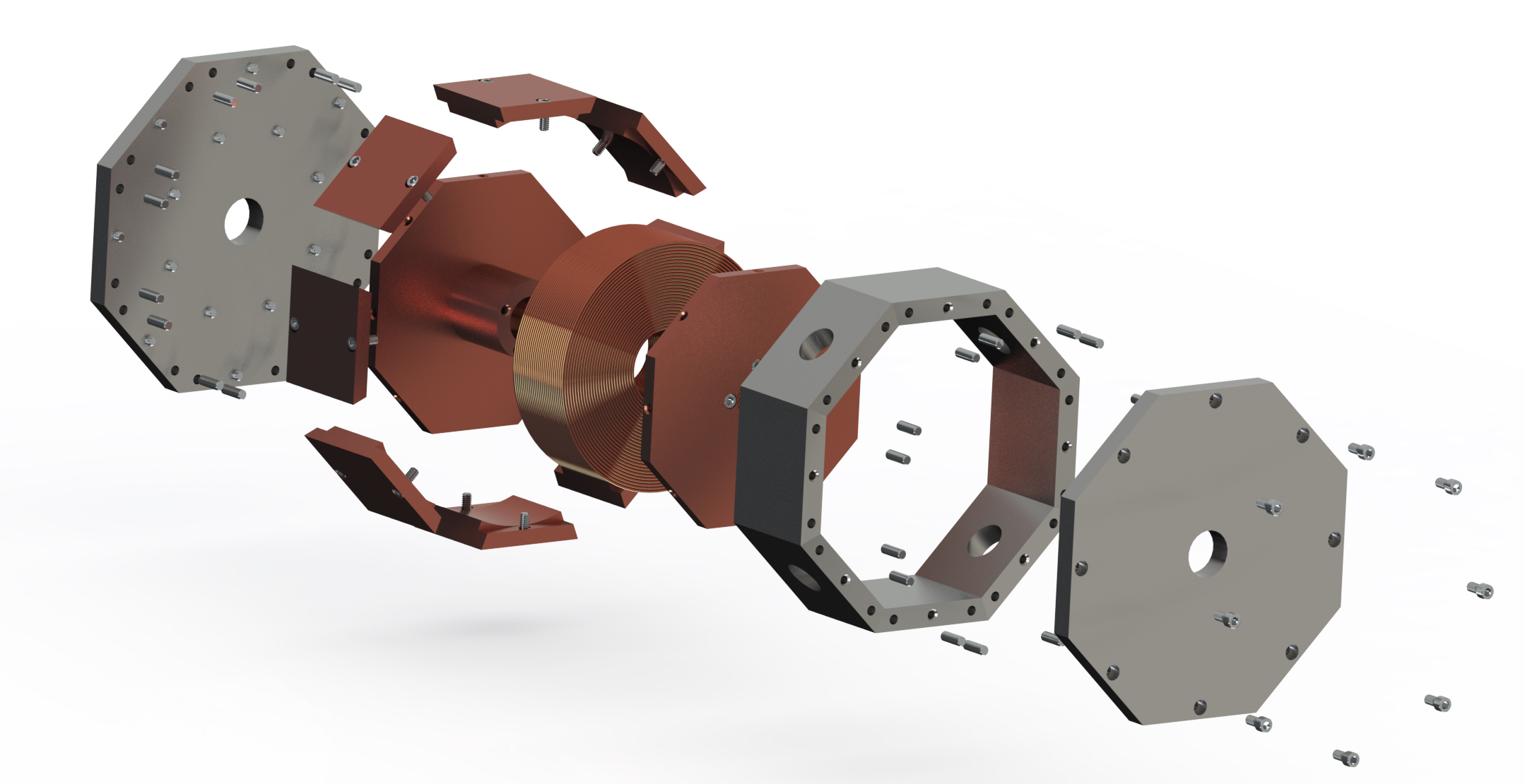}
    \caption{Render of the cryosolenoid design, showing the iron yoke, copper sarcophagus, and windings.}
    \label{fig:cryosolenoid-render}
\end{figure*}

To place a solenoid sufficiently close to the rf gun, it is necessary that it also be located inside the cryostat. It has been decided to employ a normal conducting, cryogenically cooled solenoid to avoid some of the complications associated with superconducting solenoids. However, a crucial consideration in the design of such a \textit{cryo-solenoid} is to balance the available, temperature-dependent cooling power versus the resistivity of the windings: a potentially catastrophic positive feedback loop where higher local temperatures lead to greater resistivity and thus power dissipation, overwhelming the cooling power, must be avoided (Figure \ref{fig:cryosolenoid-powerCurves}). To this end, it is necessary to choose a winding material with a low resistivity at cryogenic temperatures; this is usefully summarized by the residual resistivity ratio (RRR) which most often refers to the ratio of resistivities between 300 K and 4 K, RRR $\equiv \rho_\mathrm{300 K}/\rho_\mathrm{4 K}$. Depending on its purity, temper, and other factors, copper may exhibit RRR values from 10 to more than 5,000 \cite{simon1992properties}. 

The specific cryo-solenoid design under consideration is shown schematically in Figure \ref{fig:cryosolenoid-render}. It relies on a conventional iron yoke, RRR = 2,000 copper windings, and a copper sarcophagus for mechanical and thermal purposes: the sarcophagus serves to maintain alignment and indexing during cool down, ensure tight contact between individual wire turns, and, by virtue of being linked to the cryogenic cold head by thermal braids, serve as a heat sink for the winding Joule heating. The individual windings are to be electrically insulated with a thin film of polyimide (often referred to as \textit{Kapton}, a registered trademark of DuPont) which is well characterized at cryogenic temperatures \cite{yokoyama1995thermal}. At temperatures of interest, polyimide is approximately five orders of magnitude less thermally conductive than copper, therefore, the total thermal resistance between the heat source (windings) and sink (sarcophagus) is dominated by these polyimide layers. Electrothermal simulations incorporating temperature-dependent resistivity are conducted (Figure \ref{fig:cryosolenoid-thermal}) to determine the equilibrium thermal distribution and ensure that the runaway scenario described above does not occur. Having validated the design's thermal performance, field maps were generated using the magnetostatic code Radia \cite{chubar1998three}. The final design produces the requisite 0.51 T field, employing a current density of 9.2 A/mm$^2$, while dissipating less than 3 watts; operated at room temperature, such a solenoid would produce nearly two kilowatts. 

\section{The Role of Spatial Harmonics in rf Photoinjectors\label{sec:harmonics}}

\subsection{Description of spatial harmonic content in rf fields}

It is clear from the square-wave-like field profile that the gun we consider here (again, see Figure \ref{fig:fullcell_profile}) is rich in spatial harmonic content. In the following sections we will discuss what effects these extra harmonics have on the transverse and longitudinal beam dynamics of emittance compensation. Throughout these sections we will use a Floquet expansion of the $\pi$-mode field to guide the discussion, which defines coefficients $a_n$ and $E_0$ according to 
\begin{equation}
    E_z(z,t) = E_0 \: \text{Re} \sum_{n=-\infty}^\infty a_n\exp(i(nk_\mathrm{rf} z-\omega_\mathrm{rf}t+\phi)),
\end{equation}
and with $a_{-n}=a_{n}^*$,
\begin{equation}
    E_z(z,t) = 2E_0 \sum_{n=1}^\infty a_n\cos(nk_\mathrm{rf} z)\sin(\omega_\mathrm{rf} t+\phi).
    \label{eq:harmonics}
\end{equation}
By the indicated convention the first harmonic coefficient $a_1=1$ (see, for example, \cite{rosenzweig2003fundamentals}). With this convention, $E_0$ is the accelerating gradient observed by an ultra-relativistic particle resonant with the first spatial harmonic, having constant phase which may therefore yield maximum acceleration. Thus, in the field profile shown in Figure \ref{fig:fullcell_profile}, $E_0$ is not actually 120 MV/m (half the peak field) but rather approximately 150 MV/m. The effects of this additional acceleration have not been yet  examined in previous, very high field photoinjector studies \cite{rosenzweig2019next, rosenzweig2020ultra}. 

\begin{figure}[h!]
    \centering
    \includegraphics{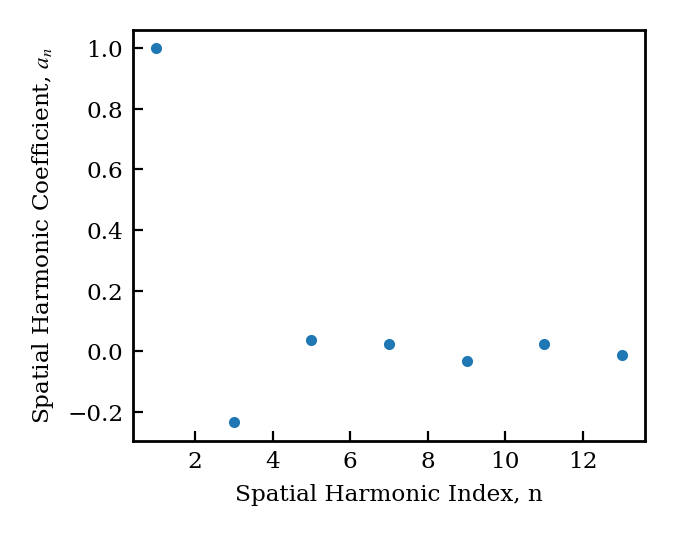}
    \caption{The Floquet coefficients of the first 13 spatial harmonics are plotted for the full cell $\pi$-mode structure.}
    \label{fig:harmonic_content}
\end{figure}

In Figure \ref{fig:harmonic_content} we plot the values of the first 13 Floquet coefficients. We observe that the primary contributions to the field profile come from the first and third spatial harmonics with $a_1=1$ by definition and $a_3\approx -0.2$. As we will see in the following sections the addition of even a single strong non-fundamental  spatial harmonic is enough to notably alter both the longitudinal and transverse beam dynamics relative to a pure single harmonic structure.

We note in this regard that that in recent decades, since the introduction of high gradient ($>$100 MV/m peak field in S-band) rf photocathode guns, the emphasis has been on use of two-cell structures having negligible higher spatial harmonic content. This approach has  been motivated by a desire to avoid potentially deleterious nonlinear field effects that can be associated with non-speed-of-light spatial harmonics. These effects are discussed below. Before engaging in this discussion, however, we examine some positive aspects of structures with higher spatial harmonic content -- the introduction of strong second order focusing effects and an enhancement of the accelerating gradient for speed-of-light particles for a fixed peak field. 

It is interesting to comment that rf structures with re-entrant nose features at the irises were indeed previously used in first-generation rf photocathode guns at LANL \cite{CARLSTEN1989313}. In these pioneering devices, such rf design features were employed to mitigate input power demands; here the same motivation exists, but it is supplemented by the possibility of obtaining strong rf-derived focusing in the gun structure, an effect augmented by the foreseen very high field operation. It is also relevant to point out that contributions to the emittance in the LANL rf photoinjectors due to nonlinear fields associated with higher spatial harmonics were significant in these designs \cite{CARLSTEN1989313}. As we will show, these nonlinear effects are greatly reduced when the beam is small enough -- a fact which is naturally realized in an ultra-high brightness injector. 

\subsection{Linear beam dynamics with spatial harmonics}

The primary transverse effect of higher spatial harmonics in standing wave accelerating structures is to introduce strong second-order ponderomotive radial focusing forces \cite{rosenzweig1994transverse}. The field-normalized strength of these forces is characterized by a parameter $\eta$ defined by 
\begin{equation}
    \eta(\phi) = \sum_{n=1}^\infty a_{n-1}^2+a_{n+1}^2-2a_{n-1}a_{n+1}\cos(\phi),
\end{equation}
where the coefficients $a_n$ measure the amplitude of the Floquet components of the on-axis axial electric field profile as we previously defined. For pure traveling wave linacs, $\eta=0$ and for pure single harmonic standing wave linacs $\eta=1$. For the gun in question, 
\begin{equation}
    \eta(\phi) = 1.12-0.5\cos(\phi).
\end{equation}
The nominal value of $\eta$ for peak acceleration is thus $1.12$, \textit{i.e.} $12$\% larger than in a pure fundamental standing wave device. Furthermore, this strength depends, with a notably large coefficient, on the phase-dependent term. In this paper we will consider primarily on-crest acceleration; however, in the rf gun itself it is not possible to inject the beam on-crest, as the velocity of the electrons changes rapidly within the gun. Further refinements of the ponderomotive theory based on the normalized velocity $\beta<1$, variations in $\eta$ and rapid relative changes in $\gamma$ may be useful, but are outside the scope of this paper.

This effective focusing increase reduces the demands placed on the focusing solenoid performance. This is a welcome development, as the requirement of using a compact, high field solenoid at relatively short rf wavelength introduces non-trivial challenges in magnet design and implementation, as discussed above.

\subsubsection{Longitudinal capture dynamics}

For high-gradient guns such as the one considered here, the beam is  accelerated from the cathode to $\beta$ near unity within the first cell. The longitudinal dynamics in this short non-relativistic portion of the gun are well-described by utilizing an effective DC field of strength $2\mu E_0\sin(\phi_0)$ \cite{serafini1995analytical}, where $\mu=\sum_{n=1}^\infty a_n$ (for our structure $\mu\approx 0.8$) and $\phi_0$ is the injection phase. During this period an accelerating electron is not perfectly resonant with the fundamental spatial harmonic. As a result, its phase slips with respect to the injection phase, and it is also accelerated by interaction with all present harmonics, giving rise to the direct sum over Floquet coefficients. The phase at which the particles exit the gun is related to that at which they are injected by the approximate relationship,
\begin{equation}
    \phi_0 = \phi-\frac{1}{2\mu\alpha\sin(\phi)}-\frac{1}{10\mu^2\alpha^2\sin^2(\phi)},
\end{equation}
where $\alpha=eE_0/k_\mathrm{rf} m_ec^2$ \cite{kim1989rf}. As Kim in \cite{kim1989rf} considers the case of a pure standing wave, he indicates the maximum electric field as $2E_0$. Here, on the other hand, the maximum field is given by $E_\mathrm{max} =2\mu E_0$. It is desirable for the beam to leave the gun with the design particle experiencing the phase of maximal acceleration, $\phi=\pi/2$, as this minimizes the transverse emittance growth associated with the spread in particle phases \cite{kim1989rf}. This yields an approximate optimal injection phase, from the standpoint of minimizing the brightness degradation from the final iris kick \footnote{This analysis is strictly valid for a 1.5 cell gun. For the 1.6 cell gun we consider here the injection phase must be offset by an additional 18$^{\circ}$ to account for the first 0.1 cell length.}, 
\begin{equation}
    \phi_0 = \frac{\pi}{2}-\frac{1}{2\mu\alpha}-\frac{1}{10\mu^2\alpha^2}.
\end{equation}
It is useful to compare this result to that of the pure standing wave case with $\mu=1$. Keeping in mind that $\mu\alpha=E_\mathrm{max}k_\mathrm{rf}/2m_ec^2$, it can be seen that the injection phase is dependent, in this approximation, on the peak field $E_\mathrm{max}$. This phase is thus not changed in this analysis by the presence of additional spatial harmonics. As in the analyses of \cite{kim1989rf} and \cite{serafini1995analytical} which assume that the slippage in phase is controlled by the field in the region directly adjacent to the cathode, this situation is indeed expected. 

The presence of additional spatial harmonics does change, however, the ratio of the final energy to the peak field. If we approximate the field in the rf gun as a square wave instead of a pure cosine form (as motivated by Figure 1), the total changes in energy due to rf acceleration over the gun length $L_g$ are: 
\begin{equation}
    \Delta U \simeq \frac{1}{2} eE_\mathrm{max}L_g \:\: (n=1 \: \text{only)}, 
\end{equation}
and 
\begin{equation}
    \Delta U \simeq \frac{2}{\pi} eE_\mathrm{max}L_g \:\: (\text{square wave}), 
\end{equation}
respectively. This result could be expected, as the ratio of the acceleration field to the peak field deduced from the spatial harmonic analysis, $E_0/E_\mathrm{max}\simeq 0.4$ for the square wave, is nearly identical to the value estimated from Eq. \ref{eq:harmonics}.

\subsubsection{Adjustments of external focusing elements}

The standard layout of a modern photoinjector consists of four basic elements. The first and most fundamental of these is the initial high-field rf gun which rapidly accelerates the beam to a few MeV off of the cathode. This is generally followed  by a short solenoid which provides a radial focusing kick to guide the beam towards a waist downstream. The third element is a drift many rf wavelengths long (meter-scale for  the C-band case considered here)  during which the beam undergoes a full transverse plasma oscillation, arriving finally at the entrance of a booster linac placed such that the beam reaches a space-charge dominated waist near the entrance to the accelerating field.

The only mandatory external focusing in this standard configuration is provided by the solenoid magnet. As we have already described, higher spatial harmonics in the field profile leads to strong ponderomotive focusing. As such, a photoinjector employing fields with spatial harmonic content may afford to have weaker solenoid focusing at what is otherwise the same operating point. Alternatively, it can operate with a smaller spot size on the photocathode as the ponderomotive focusing will mitigate excessive growth  of the beam size in the gun due to space charge defocusing. A smaller beam size at injection has the added benefit that the intrinsic thermal emittance scales linearly with the beam size on the cathode \cite{DowellSchmerge}, such that for all other parameters held constant, the photoinjector may be operated with lower intrinsic emittance. The ability to focus the beam directly as it comes off of the cathode is a unique advantage of structure-based rf focusing, as an attempt to do so using solenoidal focusing would magnetize the cathode, which is not acceptable for many commonly encountered applications.

\subsubsection{Longitudinal dynamics in downstream linacs}

In the present case the structure has been optimized to achieve 240 MV/m peak surface fields on the cathode. This is well below the maximum field achievable due to rf breakdown ($\sim$500 MV/m) \cite{CahillBreakdown}, and is limited by choice, in consideration of mitigating dark current and associated cavity power losses \cite{CahillRFloss}. As a result of the previously discussed factor $\mu$ being less than unity, the  field associated with the primary spatial harmonic is increased relative to the peak of the field profile. In the downstream linacs only this wave remains resonant with the beam and subsequently provides consistent acceleration, so the effective accelerating gradient downstream of the structure is increased relative to that of a single harmonic baseline design by 25\% for $\mu=0.8$. This is a key advantage of structures with rich spatial harmonic content: a linac structure with peak field less than its fundamental harmonic contribution can operate at a substantially larger accelerating gradient than is nominally allowed by surface-field-induced breakdown and dark current considerations. The physical explanation for this is simply that the walls of the structure see the total field at any given moment while an ultra-relativistic beam is directly sensitive only to the field of the first harmonic. This effect has a fundamental limit, at least for a $\pi$-mode cavity, which is achieved when the field profile is an exact square wave, a fact which we demonstrate in Appendix \ref{sec:fieldprof}. The enhancement factor over the peak physical field for the square wave is $4/\pi$, corresponding to an approximately 27\% increase in the gradient over what one would expect from the peak field. The similarity of this ideal figure to the 25\% enhancement in the current structure is reflective of the highly optimized nature of the cell design we consider here. 

\subsection{Nonlinear beam dynamics with spatial harmonics}

\subsubsection{Emittance dilution from nonlinear iris kicks}

The time-dependent radial momentum kick incurred at the exit iris of a 1.6 cell gun is a well-known potential source of emittance growth. In standard guns with very little spatial harmonic content the kick is perfectly linear in the radial coordinate, and the emittance growth results only from the phase spread of the electron beam in the rf wave. As we will show here, the presence of higher spatial harmonics also implies the existence of nonlinear terms in this radial kick which have the potential to dilute the beam brightness even if all particles rested at the same rf phase. 

In this section we compute the radial kick applied to the beam upon exiting the gun. We do this first for a single spatial harmonic, and then return to the Floquet expansion to compute the total kick delivered by the full field. We begin from the Floquet form of the axial electric field in the $\pi$-mode including the radial dependence \cite{rosenzweig2003fundamentals}
\begin{equation}
    E_z(\rho,z,t) = 2E_0\sum_{\text{n=1, odd}}^\infty a_n\cos(nkz)\sin(\omega t+\phi)I_0(k_{n,\rho}\rho),
\end{equation}
 where the radial wavenumber is $k_{n,\rho}^2=n^2k^2-(\omega/c)^2$ and $I_n$ is the modified Bessel function. For now, we will consider only one harmonic, $n$, and leave the summing over spatial harmonics for later. From the longitudinal electric field we can straightforwardly find the radial electric field and the azimuthal magnetic field using Maxwell's equations
\begin{equation}
\begin{split}
    E_{\rho,n}(\rho,z,t)&= 2E_0 \left(\frac{a_nnk}{k_{n,\rho}}\right)\sin(nkz)\sin(\omega t+\phi)I_1(k_{n,\rho}\rho)\\
    B_{\phi,n}(\rho,z,t)&= 2E_0 \left(\frac{a_n\omega}{c^2k_{n,\rho}}\right)\cos(nkz)\cos(\omega t+\phi)I_1(k_{n,\rho}\rho).
\end{split}
\end{equation}
The radial force on a speed-of-light particle sampling these fields is
\begin{widetext}
\begin{equation}
    F_{\rho,n}(\rho,z,t) = -e(E_{r,n}-cB_{\phi,n})= -2eE_0a_n\frac{k}{k_{n,\rho}}I_1(k_{n,\rho}\rho)(n\sin(nkz)\sin(\omega t+\phi)-\cos(nkz)\cos(\omega t+\phi)).
\end{equation}
\end{widetext}
Keeping with the traditions of similar calculations \cite{floettmann2015rf}, we extract the radial kick as a function of location in the gun in the impulse approximation, with $\rho$ held constant, assuming that $\beta=1$ after the first 0.1 cell and taking $z=0$ to coincide with the end of that first 0.1 cell region in the case of a 1.6 cell gun. Despite our explicit interest in the 1.6 cell case, this analysis holds in general for a gun of any number of cells. By integrating the radial force expression and using the fact that $t\approx z/c$, we obtain
\begin{widetext}
\begin{equation}
    \Delta p_{\rho,n}(\rho,z) = \int_0^z\frac{dz'}{c}F_{\rho,n}(\rho,z',t)= -\frac{2eE_0a_nI_1(k_{n,\rho}\rho)}{ck_{n,\rho}}[\sin(\phi)-(\sin((1-n)kz+\phi)+\sin((1+n)kz+\phi))].
\end{equation}
\end{widetext}
For the kick at the end of the 1.6 cells of the gun we evaluate this expresion at $z=3\lambda/4=3\pi/2k$. This gives
\begin{equation}
    \Delta p_{\rho,n}\left(\rho\right) =-\frac{2eE_0\sin(\phi)a_n}{k_{n,\rho}c}I_1(k_{n,\rho}\rho),
\end{equation}
where we have employed the fact that we are interested in only odd $n$. With this single harmonic kick we can evaluate the full kick via a sum over each contributing spatial harmonic,
\begin{equation}
    \Delta p_\rho(\rho) = -\frac{e}{c}2E_0\sin(\phi)\sum_n \frac{a_n}{k_{n,\rho}}I_1(k_{n,\rho}\rho).
\end{equation}
We therefore find the kick in the angular radial coordinate,  
\begin{equation}
    \Delta \rho' = \frac{\Delta p_\rho}{p_z} = -\frac{2eE_0}{\gamma mc^2}\sin(\phi)\sum_n\frac{a_n}{k_{n,\rho}}I_1(k_{n,\rho}\rho).
\end{equation}
where $\gamma$ here is in particular the beam energy at the gun exit. To extract the lowest-order correction to the common linear approximation of the radial dependence, we expand to the modified Bessel function describing this dependence, using $I_1(x)\approx (x/2)(1+x^2/8)$, to arrive at
\begin{eqnarray}
    \Delta\rho' &&\approx -\frac{2eE_0\sin(\phi)\rho}{2\gamma mc^2}\sum_n a_n\left(1+\frac{k_{n,\rho}^2\rho^2}{8}\right)\\
    &&= -\frac{2eE_0\sin(\phi)\rho}{2\gamma mc^2}\sum_n a_n\left[1+\left(n^2k^2-\frac{\omega^2}{c^2}\right)\frac{\rho^2}{8}\right]\\
    &&= -\frac{eE_z(0)\sin(\phi)\rho}{2\gamma mc^2}\left[1-\frac{\rho^2\omega^2}{8c^2}\left(1+\frac{c^2}{\omega^2}\frac{E_z''(0)}{E_z(0)}\right)\right].
\end{eqnarray}
We  retrieve the familiar linear kick in the first term \cite{floettmann2015rf}, while the second term represents a third-order kick which vanishes if there are no non-resonant spatial harmonics present. This term is suppressed by a factor $\rho^2\omega^2/8c^2$ relative to the linear term. For reference, at 1 mm off-axis in a C-band structure, this suppression is approximately three orders of magnitude.

This radial angular kick corresponds to identical $x$ and $y$ angular kicks, 
\begin{equation}
    \Delta x' = -\frac{eE_z(0)\sin(\phi)x}{2\gamma mc^2}\left[1-\frac{\rho^2\omega^2}{8c^2}\left(1+\frac{c^2}{\omega^2}\frac{E_z''(0)}{E_z(0)}\right)\right].
\end{equation}
We use this expression to estimate the emittance growth that results from the nonlinear term. For simplicity we assume the beam to be at a waist at the exit of the gun. For conciseness of notation we will write the kick in the form $\Delta x'=ax(1-\rho^2\delta)$, the relevant beam moments become
\begin{eqnarray}
    \left<x'^2\right> &&=\sigma_{x'0}^2+a^2\sigma_{x0}^2-8a^2\delta\sigma_{x0}^4+21a^2\delta^2\sigma_{x0}^6\\
    \left<xx'\right> &&=a\sigma_{x0}^2-4a\delta\sigma_{x0}^4.
\end{eqnarray}
From this result we easily find the emittance
\begin{widetext}
\begin{equation}
    \epsilon_x^2 = \sigma_x^2\sigma_{x'}^2-\sigma_{xx'}^2 = \epsilon_{x0}^2+a^2\sigma_{x0}^4-8a^2\delta\sigma_{x0}^6+21a^2\delta^2\sigma_{x0}^8-a^2\sigma_{x0}^4-16a^2\delta^2\sigma_{x0}^8+8a^2\delta\sigma_{x0}^6.
\end{equation}
\end{widetext}
In this expression we see the expected cancellation of the linear order terms, as such linear effects do not cause emittance growth, and interestingly we also see cancellation of the terms coupling the linear and nonlinear kicks. All that remains are the purely nonlinear contributions:
\begin{equation}
    \epsilon_x = \epsilon_{x0}\sqrt{1+\frac{5a^2\delta^2\sigma_{x0}^8}{\epsilon_{x0}^2}}.
\end{equation}
It is insightful to write down these terms in non-normalized variables 
\begin{equation}
    \frac{a\delta\sigma_{x0}^4}{\epsilon_{x0}} = \left(\frac{eE_z(0)\sin(\phi)}{mc^2}\right)\frac{\beta_{x0}(k\sigma_{x0})^2}{16\gamma}\left[1+\frac{c^2}{\omega^2}\frac{E''_z(0)}{E_z(0)}\right].
\label{eq:nlgrowth}
\end{equation}
The first quantity in Eq. \ref{eq:nlgrowth} is the accelerating gradient in units of electron rest energy; the second term contains the beam beta-function at the gun exit and also a term comparing the scale of the rf wavelength to the transverse beam size. The last quantity indicates the relevant nonlinear effects of spatial harmonics,
\begin{equation}
    1+\frac{c^2}{\omega^2}\frac{E''_z(0)}{E_z(0)} = 1-\frac{\sum_nn^2a_n}{\sum_na_n}.
\end{equation}
Let us evaluate this effect for the current case. The sum over harmonic contributions is roughly $1.35$. As a numerical example, we will take roughly the values corresponding to the injector design presented in Section \ref{sec:uhbbeam}: with $\sigma_x=325$ $\mu$m, $\gamma=14$, $\epsilon_{nx0}=45$ nm, and the gradient 240 MV/m, we find a roughly 1\% increase in the emittance at 45 nm corresponding to a sub-nm growth in the normalized emittance. This is nearly ignorable value, largely as a result of the relatively small beam size at the exit of the gun.  

This result is quite encouraging for the case we consider in the FEL injector section, where the beam parameters are as indicated above. However, we would eventually like to consider higher charge cases, in which the implicit scaling of the beam dimensions as $Q^{1/3}$ implies that the emittance growth term in Eq. \ref{eq:nlgrowth}, assuming  $\epsilon_{x0} \propto \sigma_x$, grows as $Q^2$. Indeed, for two cases of interest -- linear collider (discussed below) and also wakefield accelerator drivers -- we should eventually consider increasing the charge by an order of magnitude. In such cases the nonlinear contribution to the emittance from higher spatial harmonic focal effects contributes nearly equally to the final emittance as compared to more familiar effects (due to space-charge, thermal, chromatic considerations). For the moment, however, we concentrate on lower charge, higher brightness examples, which are relevant to the operation of the new C-band high gradient photoinjector initiative at UCLA that aims to enable new possibilities in both FEL and linear collider applications. 

\section{The RMS Envelope Equation for Accelerating Beams\label{sec:envelope}}

In order to prepare for the discussions of simulation results that follow, we introduce here the rms envelope equation for accelerating beams, and examine the behavior of the solutions in various relevant limits. This discussion serves the purpose of elucidating the results we will present later which are found via numerical simulations. The established approach to analyzing the dynamics of emittance compensation \cite{serafini1997envelope} utilizes longitudinal \textit{slices} of the beam, designated by the variable $\zeta=z-v_bt\simeq z-ct$. This  analysis assumes that each axially symmetric slice evolves nearly independently under the rms envelope equation, which in the limit $v_b\simeq c$ has the form
\begin{equation}
\begin{split}
\sigma_x''(z,\zeta)+\frac{\gamma'}{\gamma}\sigma_x'(z,\zeta)+\left[\frac{\eta + 2b^2}{8}\right]\left(\frac{\gamma'}{\gamma}\right)^2 \sigma_x(z,\zeta)\\ =\frac{\epsilon_{n}^2}{\gamma^2 \sigma_x(z,\zeta)^3}+\frac{I(\zeta)}{2I_0\gamma^3\sigma_x(z,\zeta)}.
\end{split}
\label{eq:rmsenv}
\end{equation}
Here we indicate the derivative with respect to the distance along the beam propagation direction with a prime symbol, $d/dz=()'$, and the constant rate of change of $\gamma$ due to interaction with the resonant (speed-of-light) spatial harmonic is $\gamma'=eE_0/m_ec^2$. Also, the local current $I(\zeta)$ is normalized to the Alfven current $I_0=ec/r_e$, the (constant under linear transformations) rms normalized emittance $\epsilon_{n}=\gamma \sqrt{\langle  x^2\rangle \langle  x'^2\rangle-\langle xx'\rangle^2}$ and $b = c B_z / E_0$. Finally, we note that the rms envelope equation self-interaction term explicitly includes only linear forces, with nonlinear effects implicitly entering only through the slice-emittance evolution.

This emittance term on the right hand side of Eq. \ref{eq:rmsenv} derives mainly, however, from the thermal-like rms spread in transverse momenta in the beam. Further, it is traditional to follow the beam dynamics governing the emittance compensation process by ignoring this term, which for relevant parameters of high brightness beams (large $I$, small $\epsilon_n$) from rf photoinjectors is small compared to the space-charge derived term. In the present work, we will examine another scenario, in which the beam is \textit{magnetized} upon emission, by placing a non-vanishing solenoidal field on the photocathode, $B_c$. In this case, one transforms the rms envelope equation to 
\begin{equation}
\begin{split}
\sigma_x''(z,\zeta)+\frac{\gamma'}{\gamma}\sigma_x'(z,\zeta)+\left[\frac{\eta + 2b^2}{8}\right]\left(\frac{\gamma'}{\gamma}\right)^2 \sigma_x(z,\zeta)\\
=\frac{\epsilon_{n}^2+L^2}{\gamma^2 \sigma_x(z,\zeta)^3}+\frac{I(z,\zeta)}{2I_0\gamma^3\sigma_x(z,\zeta)},
\end{split}
\label{eq:rmsmag}
\end{equation}
where $L=eB_c\sigma_{x,0}^2/m_e c$ is the canonical rms angular momentum owed to the cathode magnetic field. Upon leaving the solenoid region near the rf gun, this canonical angular momentum is converted into mechanical angular momentum. We will be concerned with cases where $L^2\gg \epsilon_{n}^2$, and the term proportional to $\sigma_x^{-3}$ may no longer be ignored in the envelope evolution. 

From this equation it would appear that the canonical angular momentum behaves identically to an increased thermal emittance. This is correct from the viewpoint of the envelope behavior, but hides important differences in the microscopic dynamics. In an emittance-dominated beam, the dynamics are thermal, and phase-space trajectories cross (they are nonlaminar). In the case of an angular momentum-dominated beam, however, the particle flow is laminar, and undergoes rotations.  This laminarity is a shared trait of space-charge-dominated beams. As a result, the angular momentum differs from the thermal emittance in that it is physically realized in linear correlations between the two transverse planes, which may in principle be removed by suitable downstream beamline elements.

In the following sections, we examine particular solutions of the envelope equation in the space-charge-dominated and angular momentum-dominated limits.

\subsection{Space-charge-dominated beam behavior with acceleration}

In the limit that the beam is space-charge dominated  and possesses no appreciable angular momentum, and there is no local solenoid field ($b=0$), we omit the angular momentum and emittance terms in Eq. \ref{eq:rmsenv}, and the resultant differential equation admits a particular solution, for current $I$, known as the invariant envelope \cite{serafini1997envelope}
\begin{equation}
\sigma_{x,\mathrm{inv}}(z)=\frac{2}{\gamma'}\sqrt{\frac{I}{I_0 (2+\eta ) \gamma}}.
\label{eq:invenv}
\end{equation}
The \textit{invariant} aspect of this behavior is the phase space angle of the envelope \cite{serafini1997envelope}, 
\begin{equation}
\Theta_{\sigma} \simeq \frac{\gamma\sigma'_{x,\mathrm{inv}}}{\sigma_{x,\mathrm{inv}}}=-\frac{\gamma'}{2}.
\label{eq:invenvangle}
\end{equation}
It should be noted that this corresponds exactly to the first angular kick due to entry into the linac fields \cite{rosenzweig1994transverse},
\begin{equation}
\Delta x' = -\frac{\gamma'x}{2\gamma_i},
\end{equation}
where $\gamma_i$ is the initial value of $\gamma$ in the linac.  Thus the matching of the beam to the invariant envelope in the post acceleration linac is accomplished by injecting the correct initial value ($\gamma=\gamma_{i}$ in Eq. \ref{eq:invenv}) at a waist ($\sigma'_{x}=0$). 

The emittance compensation process begins before injection into the linac, with the beam focused by a solenoid to the waist described above at the linac entrance that is similar in transverse size to that found at emission from the photocathode. At this point, the errors in trace space orientation between the various $\zeta$-slices are minimized, leaving an emittance that is a local minimum \cite{andersonNL}. It is, however, not a global minimum, as the addition of the post-acceleration linac further decreases the emittance, while adding energy to the point of the ``freezing" of the space-charge induced effects, which scale in strength as $\gamma^{-2}$. The diminishing of the emittance due to compensation in the linac is due to two effects: the further rearrangement of the slices' relative trace-space orientation; and the slow reduction of the beam size as it tends to follow the invariant envelope. In order to understand the behavior of the non-matched slices, one should examine the perturbed envelope equation for conditions close to the invariant envelope.

This analysis begins with the writing of the perturbed  envelope equation \cite{serafini1997envelope}, where we consider envelope behavior near the invariant solution, $\sigma_x(z,\zeta)=\sigma_{x,\mathrm{inv}}(z,\zeta)+\delta\sigma_x(z,\zeta)$. This can be written as
\begin{equation}
\delta \sigma_x''(z,\zeta)+\frac{\gamma'}{\gamma}\delta \sigma_x'(z,\zeta)+\left[\frac{1+\eta}{4}\left(\frac{\gamma'}{\gamma}\right)^2\right] \delta\sigma_x(z,\zeta)=0.
\label{eq:perturbedenv}
\end{equation}
We note that this differential equation is now independent of current $I$. The solution for a slice beam envelope having an error in rms size $\delta\sigma_{x,0}(\zeta)$ is 
\begin{equation}
\delta\sigma_{x}(z,\zeta)= \delta\sigma_{x,0}(\zeta)\cos \left[ \frac{\sqrt{1+\eta}}{2}\ln\left( \frac{\gamma}{\gamma_i}\right)\right],
\label{eq:perturbsol}
\end{equation}
with derivative 
\begin{equation}
\delta\sigma'_{x}(z,\zeta)= -\frac{\sqrt{1+\eta}}{2}\frac{\gamma'\gamma_i}{\gamma}\delta\sigma_{x,0}(\zeta)\sin \left[ \frac{1+\eta}{2}\ln\left( \frac{\gamma}{\gamma_i}\right)\right].
\label{eq:perturbderiv}
\end{equation}
From Eqs. \ref{eq:perturbsol} and 
\ref{eq:perturbderiv}, we can deduce that the area in trace space scales as $\gamma^{-1}$, and the phase space area is conserved. This area we term the \textit{offset emittance}, and designate it as $\epsilon_{\delta}\approx\gamma\lvert  \langle \delta\sigma_{x,0}\rangle_\zeta\langle\delta\sigma'_{x,0}\rangle_\zeta\lvert$, where $\langle.\rangle_\zeta$ indicates an average over bunch slices weighted by the current distribution. For a well-optimized design, this should be close to the thermal emittance at the cathode, perhaps increased by effects associated with rf and nonlinear space-charge.

The normalized beam emittance consists of a contribution from this offset emittance, magnified by the distance from the phase space origin to its center -- the invariant envelope. We may write an approximate expression for the behavior of the normalized emittance as 
\begin{equation}
\begin{split}
\epsilon_n & \simeq \sqrt{\epsilon_{\delta}^2+(\gamma\sigma_{x,\mathrm{inv}}\sigma'_{x,\mathrm{inv}})^2} \\ &
\simeq \sqrt{ \epsilon_{\delta}^2+\frac{4I^2}{(2+\eta)^2I_0^2\gamma'^2\gamma^2}}.
\end{split}
\label{eq:emitcompexp}
\end{equation}
This expression ignores the dynamics of slice realignment with respect to the invariant envelope trace space direction, but captures the overall behavior -- the normalized emittance approaches an asymptotic value $\epsilon_\delta$ due to the secular diminishing of the beam envelope. 

\subsection{Angular-momentum-dominated beam behavior with acceleration}

In an accelerating beam, because of the relatively strong decrease in the space-charge forces as a function of increasing energy, a beam eventually becomes emittance dominated as it attains high enough energy. In this case, one may ignore the term proportional to the peak current in Eq. \ref{eq:rmsenv}, and write a particular solution of the resulting equation as
\begin{equation}
\sigma_{\epsilon,x}=\left(\frac{8}{\eta}\right)^{1/4}\sqrt{\frac{\epsilon_n}{\gamma'}}.
\label{eq:emitdominant}
\end{equation}
Because of opposing $\gamma$-dependencies of the focusing and adiabatic damping of the emittance, the emittance-dominated solution yields simply a constant beam size. This stands in contrast to the secularly diminishing invariant envelope solution for space-charge-dominated beams. Note also that formally we may apply this result to the angular-momentum dominated case by substituting $L$ for $\epsilon_n$ in Eq. \ref{eq:emitdominant}.  

In the space-charge-dominated case, one chooses to operate at the particular solution of the envelope equation for two reasons: it is associated with a phase space angle which is independent of the local current, and it is monotonically decreasing at a rate sufficient for diminishing of the correlated emittance. This is clearly not the case in the angular-momentum-dominated case. The constant beam size associated with the particular solution in this case would not facilitate the emittance approaching the offset emittance asymptotically, rather once the beam becomes completely angular-momentum-dominated the space-charge oscillations no longer play a role in the envelope dynamics. This is highly undesirable, and as such one should inject the beam into the linac at a size large enough that the beam is still partially space-charge dominated, and in particular at a size which is notably larger than the particular solution indicated by Eq. \ref{eq:emitdominant}. This naturally prevents one from operating at the optimal spot size usually reached for a space-charge-dominated beam which is ideally close to the spot size at the cathode. This is one fundamental reason why emittance compensation with angular momentum-dominated beams is naturally less efficient. In fact, in practical scenarios, the beam spends much of its time in the injector partially space-charge-dominated, but in such a way that angular momentum cannot be ignored. 

There are yet other physics concepts at play which distinguish emittance compensation with angular momentum from that without. In fact, there are two sources which contribute to the inefficiency of emittance compensation in magnetized photoinjectors. The first is that the rms envelope equation with angular momentum has no solution which has a phase space angle independent of the local current for all energies, as we show in Appendix \ref{sec:invenvsolutions}. There is thus no clear working point analogous to the invariant envelope which achieves ideal compensation through analytical methods, rather the beam envelope must be optimized  numerically. To understand the second source of inefficiency we must consider, instead of mismatch in the phase space angle, mismatch in the local beam size. In the standard, non-magnetized case, dependence of the local slice beam size on the current contributes to what we identified already as the offset emittance, and part of the goal of numerical optimizations is to minimize this mismatch at the booster entrance. In the case of a magnetized photoinjector, reduction in the beam size mismatch is inherently less efficient. This is because in addition to the usual contributions to the emittance, a mismatch in the beam size also implies that there is no rotating frame in which every slice ceases to rotate \cite{chang2004compensation}. This results from the fact that all slices are born with the same size and thus the same canonical angular momentum, and will therefore rotate at different rates if they have different sizes further downstream. Since in practice one can never achieve perfect matching of all slice sizes, the process of minimizing the offset emittance is inherently less efficient. 

In the following sections we will give some quantitative analysis of the beam behavior in various parts of the injector and in particular discuss the ramifications of this section's discussion on the ideal injector operating point in the presence of angular momentum. 

\subsection{Beam dynamics with angular momentum in the pre-booster drift}

After leaving the compensating solenoid the beam is moderately space-charge dominated, in order that the emittance oscillations may proceed towards their eventual second minimum. In particular, for reasons elucidated in the previous section, one should ensure that the beam remains space-charge dominated until it enters the booster linac downstream. Since the angular momentum here is, by assumption, large, over-focusing could cause the beam to temporarily become angular-momentum dominated near the transverse waist, thereby interfering with the evolution of the emittance oscillations. In this section we will study the dependence of the size of the beam waist on the beam parameters upon exiting the solenoid. In this region the beam satisfies the envelope equation in the absence of acceleration and focusing,
\begin{equation}
    \sigma''-\frac{L^2}{\gamma^2\sigma^3}-\frac{I}{2I_0\gamma^3\sigma}=0.
\end{equation}
This regime is particularly difficult to deal with analytically for two reasons. First of all, as is expected, this equation does not have an analytic solution in general which is not an unwieldy infinite series. Second, although the beam is moderately space-charge dominated upon exiting the solenoid, the two terms can contribute nearly equally near the beam waist even in an optimized design, as we will see below. Nevertheless, we may determine the size of the beam at the waist by extracting a conservation law from the equation of motion. We accomplish this by multiplying the envelope equation by $\sigma'$ and interpreting the subsequent terms as exact derivatives, 
\begin{equation}
    \frac{d}{dz}\left[\frac{1}{2}(\sigma')^2+\frac{L^2}{2\gamma^2\sigma^2}-\frac{I}{2I_0\gamma^3}\log(\sigma)\right]=0.
\end{equation}
This implies that the quantity in brackets is conserved along this section of the transport. In particular then, we may relate the waist beam size $\sigma_\mathrm{min}$ to the parameters at the solenoid exit $\sigma_0$ and $\sigma'_0$ as
\begin{equation}
    \sigma_0'^2 + \frac{L^2}{\gamma^2\sigma_0^2} - \frac{2I}{2I_0\gamma^3}\log(\sigma_0)= \frac{L^2}{\gamma^2\sigma_\mathrm{min}^2} - \frac{2I}{2I_0\gamma^3}\log(\sigma_\mathrm{min}).
\end{equation}
Rewriting this in a slightly more interpretable form,
\begin{equation}
    \frac{2I_0\gamma^3\sigma_0'^2}{I} =\frac{2I_0\gamma L^2}{I \sigma_0^2}\frac{\sigma_0^2}{\sigma_\mathrm{min}^2}\left[1-\left(\frac{\sigma_\mathrm{min}}{\sigma_0}\right)^2\right]-2\log\left(\frac{\sigma_\mathrm{min}}{\sigma_0}\right).
\end{equation}
In this form we may identify the term $S\equiv 2I_0 \gamma L^2/I\sigma_0^2$ as the relative strength of the space-charge and angular momentum terms in the envelope equation at the exit of the solenoid. If we additionally define $A\equiv2I_0\gamma^3\sigma_0'^2/I$ and $x=\sigma_\mathrm{min}/\sigma_0$, this relation can be written in simpler form as
\begin{equation}
    A = \frac{S}{x^2}\left(1-x^2\right)-2\log(x) .
\end{equation}
The solution $x$ to this equation provides the size of the beam at its waist relative to the size at the exit of the solenoid. Of particular interest is the case when the beam waist corresponds to the beam size at which the angular-momentum and space-charge terms become equivalent, as past this point one can consider the angular-momentum term dominant, thereby interfering with the emittance compensation process. Under this condition, $x_0=(L/\sigma_0)\sqrt{2I_0\gamma/I}=\sqrt{S}$. Thus $A=1-S-2\log(\sqrt{S})$. If $A$ exceeds this value, the beam over-focuses and becomes angular momentum dominated at the waist. As a result, one should design the injector such that $A$ is below this cutoff. In the simulations studies we present below, we find through numerical optimization that in fact the ideal scenario is to operate such that $x_{\min}\approx\sqrt{S}$. This is unsurprising: in general the emittance compensation process is more efficient with smaller beam size and this represents the smallest beam size that allows emittance compensation to proceed without interruption.

\subsection{Beam behavior in the booster linac of a magnetized photoinjector}

In a traditional high-brightness (unmagnetized) injector the aim of the booster linac is to facilitate the transition of the beam from space-charge dominance to emittance dominance without interrupting the final stage of emittance compensation, in particular by optimally damping the transverse beam size with increasing energy. This increase in energy serves to diminish the relative strength of the space-charge forces. To optimally make the transition to emittance domination, one operates at or near the invariant envelope, where the beam size drops as $\sigma_x\propto1/\sqrt{\gamma}$, and the dominant slice dynamics are independent of the local beam current, thereby achieving the goal of allowing the emittance compensation dynamics to proceed to their eventual minimum-emittance configuration. As we demonstrated in previous sections, such a solution does not exist in the presence of notable emittance or angular momentum effects. Furthermore, the predominance of angular momentum in the magnetized beam case implies that the described transition occurs at a much lower energy, thereby occurring earlier for equal accelerating gradients. In order to operate at moderate gradients while arresting the emittance evolution near the minimum of the emittance oscillations, this implies placing the booster linac slightly further downstream than the transverse beam waist -- consistent with the qualitative analysis of \cite{chang2004compensation}. Furthermore, to facilitate the diminishing of the emittance we would like to employ an alternative approach to reducing the beam size, inspired by the $1/\sqrt{\gamma}$ dependence of the invariant envelope. This indeed implies operation far from the equilibrium beam size associated with the pure angular-momentum-dominated solution, thus an accurate analytic description of the envelope progression for ideal compensation in this case demands a relatively elaborate approximation scheme. 

Here we present a simple model for the beam envelope evolution in the relevant regime, where the beam enters the booster under space-charge dominated conditions, but  becomes angular momentum dominated rapidly thereafter. This analysis is intended to elucidate the dominant processes involved in the envelope evolution in the booster linac of an injector with angular momentum. It will also serve to provide a method for describing how the beam transitions from a space-charge-dominated state to an angular-momentum-dominated one.

\begin{figure}[h!]
    \centering
    \includegraphics{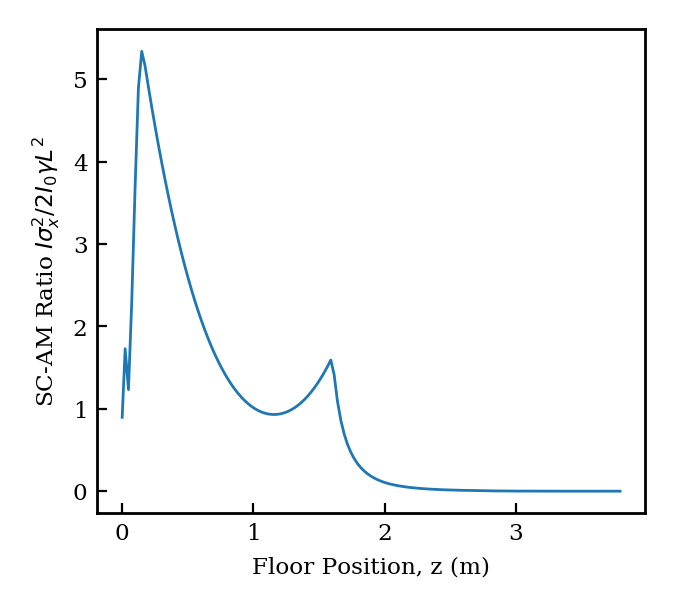}
    \caption{The ratio of the space-charge and angular momentum terms in the envelope equation is plotted along the injector studied in Section \ref{sec:magbeam}.}
    \label{fig:sc-am}
\end{figure}

At entrance into the booster linac the space-charge contribution to the envelope equation only slightly exceeds that from angular momentum. This situation is described clearly by Figure \ref{fig:sc-am} where we have plotted the ratio of the space-charge and angular momentum terms in the envelope equation through the injector design presented in more detail in Section \ref{sec:magbeam}. As discussed above, at the beam waist the space-charge and angular momentum terms contribute nearly equally to the envelope oscillations, and the optimal solution requires allowing the beam to expand to a state in which it is more heavily space-charge dominated before injecting it into the linac. When the beam does enter the booster at roughly $z=1.6$ m, the transition to angular-momentum-dominance is rapid due to the effects of acceleration and occurs before the beam size changes appreciably. As such, we anticipate that the contribution of space-charge to the beam dynamics may be sufficiently captured by approximating the space-charge term in the envelope equation by evaluating it only employing the beam size at linac entrance, 
\begin{equation}
    \sigma''+\frac{\gamma'}{\gamma}\sigma'+\frac{\eta}{8}\left(\frac{\gamma'}{\gamma}\right)^2\sigma = \frac{L^2}{\gamma^2\sigma^3} + \frac{\kappa}{\gamma^3\sigma_0}.
\label{eq:magenvelope}
\end{equation}
Since we assume the dynamics along the majority of the length of the linac will be dominated by the angular momentum contribution, we will separate the envelope solution into two terms: $\sigma(z)=\sigma_\mathrm{AM}(z)+\sigma_\mathrm{SC}(z)$. We will consider space-charge as a perturbation such that $\sigma_\mathrm{AM}(z)$ satisfies
\begin{equation}
    \sigma_\mathrm{AM}''+\frac{\gamma'}{\gamma}\sigma_\mathrm{AM}'+\frac{\eta}{8}\left(\frac{\gamma'}{\gamma}\right)^2\sigma_\mathrm{AM} = \frac{L^2}{\gamma^2\sigma_\mathrm{AM}^3}
\end{equation}
This equation has an exact solution, given by \cite{floettmann2017emittance}
\begin{equation}
\begin{split}
    \sigma_\mathrm{AM}^2(z) =& \sigma_i^2\cos^2(\psi(z))+\frac{4\sqrt{2}\sigma_i\gamma_i\sigma'_i}{\gamma'\sqrt{\eta}}\sin(\psi(z))\cos(\psi(z))\\
    &+\frac{8}{{\gamma'}^2\eta}\left(\frac{L^2}{\sigma_i^2}+\gamma_i^2{\sigma'_i}^2\right)\sin^2(\psi(z)),
\end{split}
\label{eq:angmosolution}
\end{equation}
where 
\begin{equation}
    \psi(z) = \sqrt{\frac{\eta}{8}}\log\left(\frac{\gamma}{\gamma_i}\right).
\end{equation}
Here, in the absence of space-charge forces, $\sigma_i$ and $\sigma'_i$ would be the transverse size and angle at the start of the linac, but for now they should be thought of as arbitrary values which we will fix after describing the space-charge contribution. 

Next, we show that there is a more accurate way to approximate the dynamics, by modifying these initial conditions in the angular momentum envelope beyond inclusion of the space-charge term in the constant beam size approximation. Returning to this space-charge term, we determine the evolution of $\sigma_\mathrm{SC}$ by linearly expanding Eq. \ref{eq:magenvelope} about $\sigma_\mathrm{AM}(z)$ while neglecting the residual angular momentum term, a choice which we justify retroactively by demonstrating the validity of these approximations through numerical integration of the envelope equation. With this choice, we now write 
\begin{equation}
    \sigma_\mathrm{SC}''+\frac{\gamma'}{\gamma}\sigma_\mathrm{SC}'+\frac{\eta}{8}\left(\frac{\gamma'}{\gamma}\right)^2\sigma_\mathrm{SC} = \frac{I}{2I_0\gamma^3\sigma_0}.
\end{equation}
The particular solution to this equation is 
\begin{equation}
    \sigma_\mathrm{SC}(z) = \frac{4I}{I_0\gamma'^2\gamma(8+\eta)\sigma_0}
\label{eq:scenvelope}
\end{equation}
We further note that this does not have $\sigma_\mathrm{SC}(0)=0$ and $\sigma'_\mathrm{SC}(0)=0$, thus if we assume $\sigma_i=\sigma_0$ and $\sigma'_i=\sigma'_0$ in Eq. \ref{eq:angmosolution} we would be contradicting our initial conditions. We can resolve this in one of two ways: either including the homogeneous solution in the space-charge solution and choosing its coefficients accordingly, or by taking, in Eq. \ref{eq:angmosolution},
\begin{equation}
\begin{split}
    \sigma_i &= \sigma_0 - \frac{4I}{I_0\gamma'^2\gamma_i(8+\eta)\sigma_0}\\
    \sigma'_i &= \sigma'_0 + \frac{4I}{I_0\gamma'\gamma_i^2(8+\eta)\sigma_0}.
\end{split}
\end{equation}
With this choice our initial conditions are consistent for the total beam envelope $\sigma(z)=\sigma_{AM}(z)+\sigma_{SC}(z)$. 

\begin{figure}[h!]
    \centering
    \includegraphics{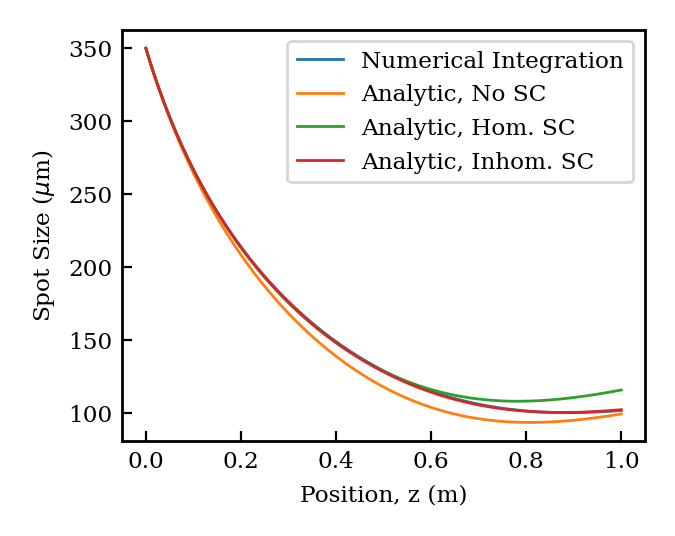}
    \caption{The beam envelope in the booster linac presented in Section \ref{sec:magbeam} is determined four different ways, with each described in the text.}
    \label{fig:envelope_comparison}
\end{figure}

The seemingly arbitrary choices made in developing this derivation demand some justification, which we provide in Figure \ref{fig:envelope_comparison}. In this figure the beam and accelerator parameters used correspond to those for the first booster linac presented in Section \ref{sec:magbeam}. In it we have plotted four different solutions to the envelope equation. The first, the blue line which is concealed by the red line, is obtained by numerically integrating the envelope equation with no approximations or assumptions. The second, labeled ``Analytic, No SC" is simply Eq. \ref{eq:angmosolution} with $\sigma_i=\sigma_0$ and $\sigma_i'=\sigma_0'$. The third, labeled ``Analytic, Hom. SC" is the sum of the analytic angular momentum and space-charge envelopes with $\sigma_i=\sigma_0$ and $\sigma_i'=\sigma_0'$ and the homogeneous terms included in Eq. \ref{eq:scenvelope} to make $\sigma_\mathrm{SC}(0)=0$ and $\sigma'_\mathrm{SC}(0)=0$. The final line, labeled ``Analytic, Inhom. SC" is the solution we have presented as optimal in the derivation. This claim appears to be fully justified by the figure, where this solution lies on top of the numerical integration. The analytic solution ignoring space-charge finds a transverse waist which is both too small and occurs too early due to the complete absence of the space-charge defocusing force. The solution with homogeneous terms maintained in the space-charge contribution approaches a waist too early and with too large of a minimum value. 

Unlike the result of the previous section, this analysis does not rely on the beam entering the linac near the constant envelope solution associated with the angular-momentum-dominated dynamics. For the reasons stated above, this is much more relevant to the practical optimization of an injector with a high degree of angular momentum.

\section{\label{sec:uhbbeam}Ultra-High Brightness Mode for an XFEL Driver}

\subsection{Beam requirements for an advanced XFEL}

We next concentrate on a highly relevant example of high brightness, unmagnetized beam production in the high field photoinjector, which is applied to a frontier x-ray free electron laser (XFEL) scenario. In this regard, recent work towards the design and realization of an ultra-compact XFEL (UC-XFEL) with a footprint on the order of 40 m has highlighted the necessity of producing beams with exceptional 6D brightness \cite{rosenzweig2020ultra}. The existence of such a device is largely predicated on the ability to use ultra-high gradient accelerating structures to produce a GeV-scale beam in only 10-20 meters. The linac design required to achieve this goal has largely been demonstrated \cite{Tantawi2020}, and further studies aimed at realizing mature devices based on this technology are currently being pursued by a SLAC-UCLA-LANL-INFN collaboration \cite{rosenzweig2020ultra}. Nevertheless, scaling the energy of the electron beam down to 1 GeV while fixing the x-ray photon energy demands a notable, ambitious decrease in the electron beam normalized emittance to enable robust lasing. Thus, the state-of-the-art status of free-electron laser photoinjectors, producing some 10's of ampere beams with 0.2 $\mu$m rad scale normalized emittances, is not adequate to realize an ultra-compact XFEL on the desired scale. 

In previous work \cite{rosenzweig2018ultra,rosenzweig2019next} it was determined via simulations employing ideal rf and solenoid field maps that with a 240 MV/m peak field on a photocathode, an electron beam of $~55$ nm rad normalized emittance and 17 A current could be produced. It has additionally been demonstrated in simulation that this exceptional brightness can be preserved with minimal degradation during transport, acceleration (to 1 GeV) and longitudinal compression \cite{robles2019compression, rosenzweig2020ultra} by a factor of 200. The ultra-high brightness operating regime for the presently considered photoinjector serves to explore and validate the ability to produce beams of similar quality to those proposed in previous iterations of the gun when a more complete view of the rf and solenoid fields used is employed. Further, we deepen the previous studies by including a simulation study of an important effect in such unprecedented \textit{6D} brightness systems, that of intra-beam scattering (IBS) \cite{Di_Mitri_2020}. This effect, which until now has not been examined by detailed simulations of the beam's microscopic behavior, may have notable negative implications for 6D phase space dilution -- introduction of unwanted additional slice energy spread.   

It is also worth noting that constraints similar to those of the UC-XFEL regarding beam brightness can be found in large-scale XFEL applications demanding high photon flux at high photon energy, such as the proposed MaRIE XFEL. A design study of this particular potential machine found that a beam source of the quality we seek to demonstrate here would substantially decrease the difficulties associated with reaching these extreme performance metrics \cite{carlsten2019high}. 

\subsection{Comment on thermal emittance}

In order to understand the low emittance presented in the simulations to follow, a discussion of the factors which enable such a low emittance is demanded. This discussion naturally begins with the thermal emittance at the cathode. We recall that this has the form \cite{DowellSchmerge} 
\begin{equation}
    \epsilon_{n,th} = \sigma_x\sqrt{\frac{\mathrm{MTE}}{3mc^2}}
\end{equation}
where the mean transverse energy MTE is defined by $\mathrm{MTE}=\hbar\omega-\phi_{\mathrm{eff}}$ where $\hbar\omega$ is the photon energy of the cathode drive laser and $\phi_{\mathrm{eff}}$ is the effective cathode work function after the Schottky effect has been accounted for \cite{schottky1918spontane}. In the present scenario we envision an MTE of 140 meV, consistent with a cathode drive laser of 262.1 nm in this simple model. This results in a thermal emittance, quoted in the usual way, of 0.3 $\mu$m/mm. This value, although slightly ambitious, is not far from the state-of-the-art for copper cathodes in photoinjectors \cite{divall2015intrinsic}, and is well above that achieved at test stands \cite{Karkare2020}.  Further, it has been proposed that notable improvements may be made with cryogenic operation of the injector and thus the photocathode \cite{cry-cathode}. This and other approaches promise to lower the thermal emittance in the near future, and the subject of exploiting much smaller MTE is actively being explored at present \cite{IntrinsicEmit}.

The use of low thermal emittances to enable the high brightness we seek here implies a small beam size on the cathode of roughly 100 $\mu$m, assuming adequate quantum efficiency. This question of the impact of and questions surrounding quantum efficiency will be discussed within the experimental context in the conclusion. This small spot size injection is enabled by several features of the injector design presented already: namely strong ponderomotive rf focusing and rapid acceleration in a high-field environment. Furthermore, since the emittance is already so small, any unexpected increases to the thermal emittance would roughly add in squares to the optimal value achieved in simulations below and would be unlikely to require much in the way of changes to the injector operating point we consider. This is expected to be the case as long as emittance does not grow large enough so as to jeopardize the space-charge-dominated nature of the envelope evolution. Thus even with more standard thermal emittance figures we would not expect the final emittance to exceed 55 nm rad, which is the value found in previous iterations of this very gun design. The reason for this is that the final emittance is determined roughly in equal parts by thermal emittance considerations and other factors such as rf and nonlinear space-charge emittance growth.

\subsection{Photoinjector performance}

\begin{figure}[h!]
    \centering
    \includegraphics{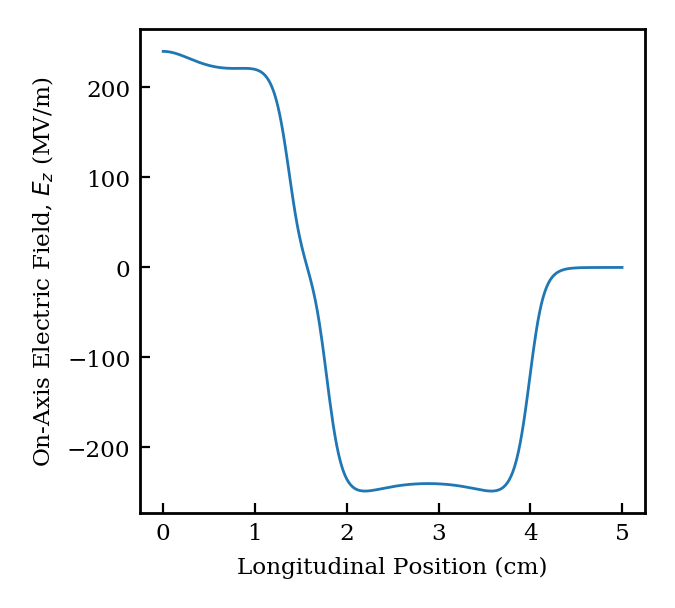}
    \caption{The on-axis field profile of the 1.6 cell gun design is shown. }
    \label{fig:gunprofile}
\end{figure}

The simulations for both the high-brightness operating conditions and those relevant to flat beam (magnetized photocathode) operation are performed using the General Particle Tracer (GPT) code with 350k macroparticles with the ``accuracy" parameter set to six \cite{GPTSite}. In these initial studies we  employ a mesh-based three-dimensional space-charge routine suitable for capturing the complex collective physics in the gun but which elide over microscopic space-charge effects such as IBS and disorder-induced heating (DIH) \cite{Maxson_2013}. These effects are discussed in detail in later sections. For both operating modes, the beam distribution is uniform in the longitudinal coordinate and a gaussian cut off at $1\sigma_r$ in the transverse coordinates. The high-brightness operating mode of the injector consists of the high-gradient 1.6 cell gun, whose complete on-axis field profile in shown in Figure \ref{fig:gunprofile}, a cryogenically cooled compensating solenoid, and a 40 cell C-band booster linac structure of roughly one meter length \cite{Tantawi2020}.

\begin{table}[h!]
    \centering
    \begin{tabular}{|c|c|c|}
    \hline
        Parameter & Unit & Value   \\
    \hline
        Charge & pC & 100\\
        Laser Spot Size (Pre-Cut) & $\mu$m & 151\\
        Laser Spot Size (Post-Cut) & $\mu$m & 76\\
        Injection Phase & $^{\circ}$&44  \\
        Laser Length & ps & 5.8\\
        Peak Cathode Field & MV/m & 240\\
        \hline
        Solenoid Field & T & 0.51\\
        Solenoid FWHM & cm & 7.4\\
        Solenoid Center & cm & 12.5\\
        Booster Gradient & MV/m & 77\\
        Booster Entrance & m & 1.165\\
        Booster Phase & $^{\circ}$ & 90\\
        \hline
    \end{tabular}
    \caption{The injector parameters relevant to emittance compensation for ultra-high brightness operation are listed. }
    \label{tab:uhbparameters}
\end{table}

\begin{figure}[h!]
    \centering
    \includegraphics{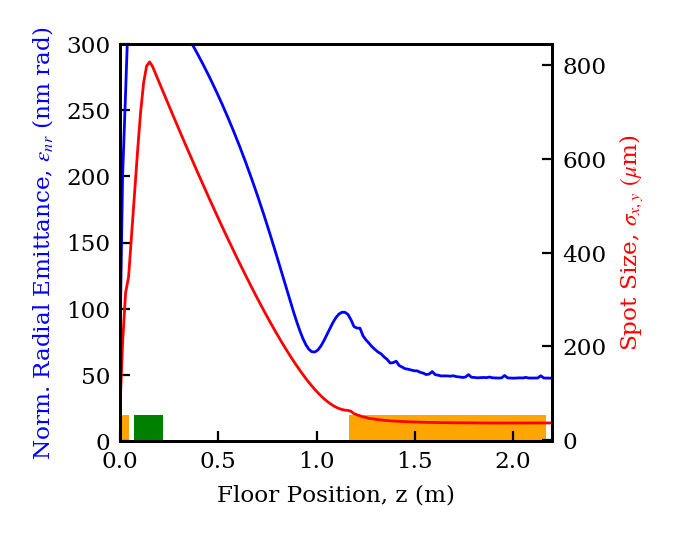}
    \caption{The emittance and beam size evolution of the ultra-high brightness operating point are shown through the first booster linac.}
    \label{fig:uhbemittance}
\end{figure}

The ultra-high brightness working point has been optimized to generate the smallest possible normalized emittance with just under 20 A peak current at 100 pC bunch charge, as was found in previous, less mature iterations of the gun design. The final parameter set, found through numerical minimization of the beam emittance, is given in Table \ref{tab:uhbparameters}. The result of this optimization is found in Figure \ref{fig:uhbemittance}. Here we show the normalized  emittance in blue alongside the root-mean-square spot size in red as the beam travels through the injector. In addition, we include graphics displaying the locations of the gun, the solenoid, and the 1 m-long booster linac. We note that in these simulations, we have employed the spatial harmonic-rich structure described in detail above for both the gun and booster linac. 

The emittance compensation profile evolution is relatively standard, with injection into the linac occurring at the usual ``Ferrario working point" at which the beam enters the booster at its waist. Further, the beam is very nearly on the  invariant envelope corresponding to the parameters of this high gradient standing wave linac. This invariant envelope initiates at a $~64$ $\mu$m spot size waist according to Eq. \ref{eq:invenv}, which is nearly exactly the $63$ $\mu$m waist found through numerical optimization. As a result we observe the expected simultaneous damping of both the emittance and the beam size, as described qualitatively above -- the emittance has a value of roughly 45 nm rad as it exits the booster linac, a 10 nm rad improvement over previous iterations of the gun. This enhanced performance may be attributed to the beneficial effects associated with spatial harmonic content in the gun and linac. Furthermore, the transition to emittance-dominated beam transport is almost ideally accomplished in this first booster linac, as the beam finds an asymptotic value of 37 $\mu$m, which is almost exactly equal to the quadrature sum of the invariant envelope and the emittance dominated solution $\sigma_{\epsilon,x}=\sqrt{\sqrt{(8/\eta)}\epsilon_n/\gamma'}$.

We also highlight the dramatic expansion of the transverse distribution through the gun and solenoid, at the peak of which the beam has grown relative to its size at the cathode by roughly an order of magnitude. This is consistent with the findings of the authors of Ref. \cite{andersonNL}, who showed that emittance compensation is most effective when the beam propagates far from equilibrium. In this way, local phase space wave-breaking due to nonlinear field effects is avoided. This rapid expansion also serves to rearrange the beam distribution to produce more linear self-fields, through the transverse "blowout" effect. 

\begin{figure}[h!]
    \centering
    \includegraphics{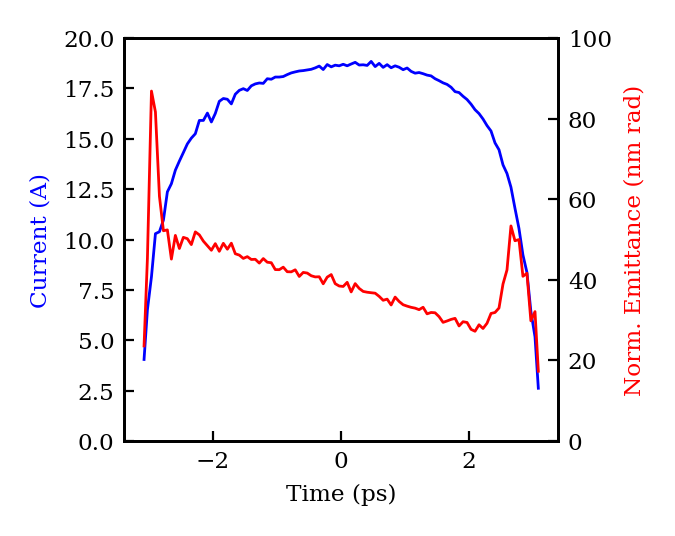}
    \caption{The time-dependence of the current (blue) and normalized transverse emittance (red) are plotted for the FEL injector case. }
    \label{fig:uhbslices}
\end{figure}

We next explore the beam distribution at the linac output in Figure \ref{fig:uhbslices}. Here we plot in blue figure the current profile as a function of longitudinal position in the bunch, referenced by the time-of-flight parameter, and in red the normalized slice emittance. We see that the beam current is very nearly uniform with a peak of 19 A. The slice emittance is seen to be relatively uniform itself, with an average value in the core nearly equal to the value of the projected emittance at the exit of the booster linac. This feature is indicative of a highly optimized emittance compensation process. The only slice parameter of interest which is not plotted is the slice energy spread, which has a roughly uniform value of 300 eV across the bunch length. The small value of the energy spread as compared to state-of-the-art FEL injectors can be attributed to two features of our design: the relatively high accelerating gradient, and the C-band design frequency which naturally allows less time for space-charge-induced slice energy spread to develop. 

\begin{figure}[h!]
    \centering
    \includegraphics{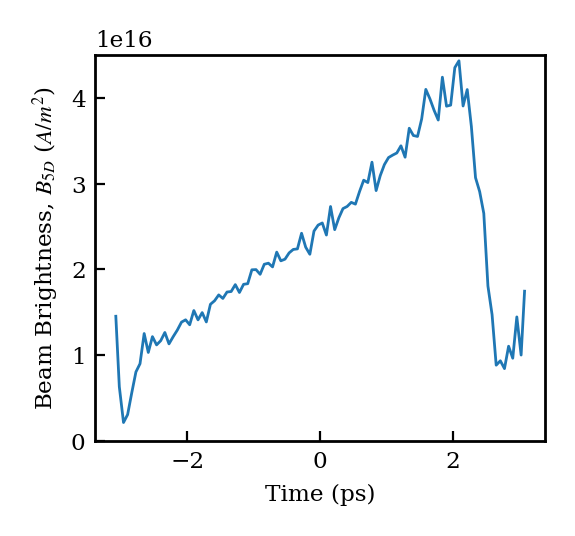}
    \caption{The time-dependence of the five-dimensional brightness, $B_{\mathrm{5D}}=2I/\epsilon_{nx}^2$, is plotted for the FEL injector case. }
    \label{fig:uhb5dbright}
\end{figure}

These results  also present the opportunity to discuss the five-dimensional beam brightness, defined as $B_{\mathrm{5D}}=2I/\epsilon_n^2$ with $I$ the current and $\epsilon_n$ the normalized emittance. We show the slice-dependence of the five-dimensional brightness in Figure \ref{fig:uhb5dbright}, where it is seen to take an average value of $2\times 10^{16}$ A/m$^2$. This should be compared to the state-of-the-art photoinjectors driving modern XFELs, such as the present SwissFEL injector where the 20 A current and 200 nm rad emittance yields $B_{\mathrm{5D}}\simeq 10^{15}$ $A/m^2$ \cite{prat2019generation}. This is a dramatic increase in brightness, an advance that has few comparisons in the recent history of high  brightness electron beam production.

\section{\label{sec:microscopic}Microscopic Space-Charge Effects}

\subsection{Short-range coulomb interactions in photoinjectors}

The injector design as presented thus far produces, in addition to an extremely low emittance, an energy spread which is well below the state-of-the art. Where a standard photoinjector produces a beam of several keV energy spread, we have presented here a beam with a final slice energy spread of just a few hundred eV. While in a traditional FEL architecture this difference has relatively small impact, as the use of a laser heater downstream would purposefully increase the energy spread,  we now find applications which could not only make use of this exceedingly low energy spread but may also require it. Indeed, the novel UC-XFEL design presented in \cite{rosenzweig2020ultra}, does not utilize a laser heater and in fact has an upper bound on input energy spread of ~10 keV. Additionally, such a source, if operated at lower charge, would be interesting for ultra-fast electron diffraction and microscopy where the uncorrelated energy spread presents a limit on achievable resolution \cite{li2014single}. 

In part because the predicted low energy spread of this source is quite promising, it is also subject to high levels of scrutiny concerning the validity of the space-charge interaction model employed in the simulations up to this point. Mesh-based space-charge algorithms inherently elide over the \textit{microscopic} physics associated with short-range Coulomb interactions which can give rise to phase space dilution in two different forms. The first, intra-beam scattering (IBS), is a result of the short-range predominantly binary scattering events between neighboring electrons which can lead to phase space dilution, primarily through an increase in the uncorrelated energy spread of the beam \cite{Di_Mitri_2020}. The second, disorder-induced heating (DIH), results from an initially spatially disordered beam distribution seeking out a pseudo-crystalline configuration of lower potential energy, similarly mediated by short-range Coulomb interactions \cite{Maxson_2013}.  In lowering the potential energy of its charge configuration,  the beam increases its uncorrelated kinetic energy, a fast process (time-scale $\omega_p^{-1}$) by which an increase in both the emittance and the energy spread is realized. These \textit{granular} effects demand a space-charge algorithm capable of taking into account point-to-point space-charge effects, at least for electrons which are near each other, while avoiding large numerical errors from close encounters. This may be realized in a full point-to-point space-charge calculation, which is computationally unwieldy, or more practically with a Barnes-Hut approach \cite{Barnes1986AHO}.

Of these two effects, it is expected that IBS presents a more important limit to the applicability of the high field photoinjector operated with 100 pC charge since it is expected to yield a contribution to the uncorrelated energy spread on the order of a keV for the beam charge considered \cite{rosenzweig2020ultra}.  While present injectors can be evaluated acceptably by either ignoring or estimating the approximate IBS energy spread, as was done in both \cite{qiang2017start} and \cite{rosenzweig2020ultra}, in the present design the IBS contribution may be comparable to or even larger than the non-IBS spread. DIH, on the other hand, has the primary effect of increasing the beam emittance and is typically much smaller than the thermal emittance employed in the 100 pC gun. It can, however, become relatively more important in a design scaled down to the much lower charges relevant to UED and UEM, since the thermal emittance scales linearly  with the beam size on the cathode, or the charge as $Q^{1/3}$, while the space-charge contribution scales as  $Q^{2/3}$. 

Analyzing both of these effects is computationally unwieldy for the 100 pC operating point we have discussed so far, with its associated 625 million electrons. However, both effects are expected to scale with the beam density $N_e/\sigma_x^2\sigma_z$, so we can quantify the IBS-induced energy spread in the 100 pC gun by scaling all beam dimensions down an order of magnitude while dropping the charge by three orders of magnitude to 100 fC. This procedure preserves the space-charge dominated beam dynamics of the beam, including emittance compensation, in the relevant low-energy region.  The resulting beam has 625,000 electrons, a number which is feasible to simulate with a Barnes-Hut space-charge algorithm and which is thus able to capture the essential features of short-range Coulomb interactions. 

\subsection{Scaled simulations of IBS energy spread}

We have performed the aforementioned scaled simulations of the ultra-high brightness injector by reducing the charge by three orders of magnitude and all physical beam dimensions by one order of magnitude so as to preserve the beam charge density which determines both the emittance compensation dynamics and the strength of the short-range Coulomb interactions. The reduction in the transverse size at emission implies a concomitant reduction in the thermal emittance, which changes linearly with the spot size on the cathode. In fact, the translation of the 100 pC case to the 100 fC case is not quite exact because of certain effects which do not scale with the beam density: in particular the contribution to the emittance owed to time-dependent rf effects and the space-charge field of the image charges in the cathode. As a result, the emittance compensation profile is not perfectly retrieved at low charge without modifying other parameters in the injector. Nevertheless, since we are primarily interested in the effect on the energy spread and would like to maintain a proper comparison between the two cases, we will neglect the effects on the emittance compensation and consider the energy spread growth which occurs only up to the entrance to the booster linac. This is allowable because the dominant source of energy spread growth occurs when the beam is densest -- at the waist ahead of the booster -- and thus by stopping the simulation at this point we will have captured the majority of the IBS contribution in the injector. The scaled simulations are also performed using GPT. In order to properly account for the cathode image charge effect, three space-charge routines are employed simultaneously: the Barnes-Hut simulation which does not account for image charges in the cathode, a mesh algorithm including the image charges, and a mesh algorithm not including the image charges with an overall minus sign applied to the space-charge field. In this way, one can take into account the image charge field using a mesh algorithm and the real beam space-charge field using a Barnes-Hut algorithm without redundant application of the beam's own space-charge field. This is the same procedure described in the GPT manual \cite{van2009general}.

\subsubsection{Scaled simulation results}

\begin{figure}[h!]
    \centering
    \includegraphics{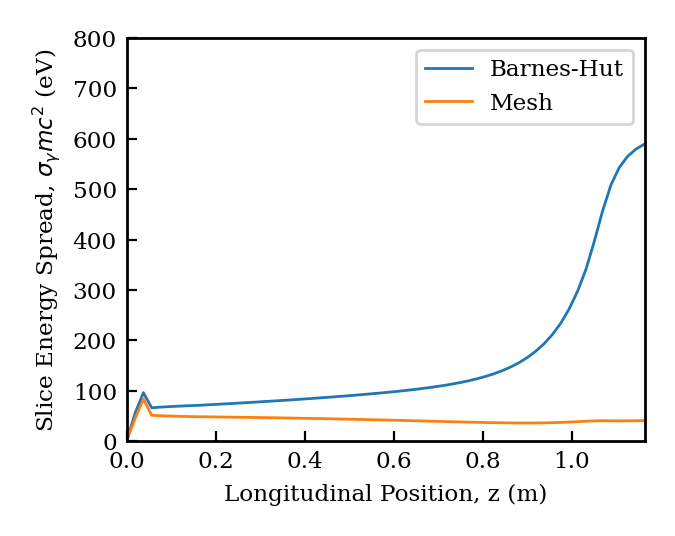}
    \caption{The uncorrelated energy spread is shown from two different simulations of a scaled 100 fC injector. The Barnes-Hut simulation (blue) demonstrates the energy spread growth from short-range Coulomb effects relative to a baseline case (orange) which utilizes a mesh-based space-charge algorithm. }
    \label{fig:spread_ibs}
\end{figure}

We have run this scaled simulation and extracted the uncorrelated energy spread of the beam throughout the gun and drift prior to the booster linac, which we display in Figure \ref{fig:spread_ibs} alongside similar data from another scaled simulation using a mesh-based space-charge algorithm. We observe that, as expected, the energy spread is increased by the short-range Coulomb interactions included in the Barnes-Hut simulation. In particular, the corresponding energy spread growth is most dramatic as the beam approaches a transverse waist at the entrance to the booster linac at $z=1.165$ m, as this is the point where the beam density is at its largest value. In these simulations, as we will see in Figure \ref{fig:ibs_validation}, this waist actually appears closer to $z=1.05$ m, hence the peak in the gradient of the uncorrelated energy spread occurring nearer to this point. When the full booster linac is included, the energy spread only increases by roughly an additional 100 eV. 

\begin{figure}[h!]
    \centering
    \includegraphics{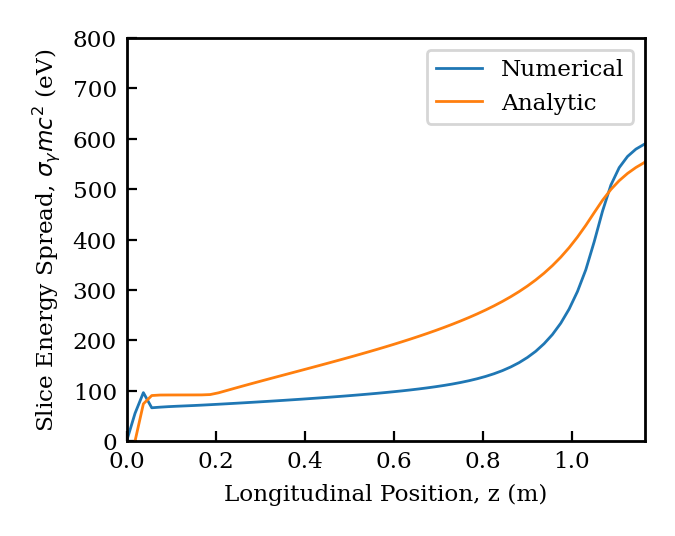}
    \caption{The uncorrelated energy spread is shown as computed by the analytic formula (orange) and as produced by the Barnes-Hut simulation (blue). }
    \label{fig:spread_comparison}
\end{figure}

We may attempt to contextualize this energy spread growth by comparing it to the theoretical predictions made in \cite{huang2002intrabeam}. In that work, it was found that the energy spread growth rate satisfied 
\begin{equation}
    \frac{1}{\sigma_\gamma}\frac{d\sigma_\gamma}{dz} = \frac{r_e^2N_b}{\sigma_x\sigma_z\epsilon_{nx}\sigma_\gamma^2},
\end{equation}
where $r_e=e^2/4\pi\epsilon_0mc^2$ is the classical electron radius and $N_b$ is the number of particles in the bunch. The original paper \cite{huang2002intrabeam} treated the rms beam quantities which appear in this expression in an averaged sense over the length of the injector, however one can just as easily include their z dependence. This may be more conveniently written in the form of the derivative of the square of the energy spread,
\begin{equation}\label{eq:ibs}
    \frac{d\sigma_\gamma^2}{dz} = \frac{2r_e^2N_b}{\sigma_x\sigma_z\epsilon_{nx}}.
\end{equation}
In Figure \ref{fig:spread_comparison} we have plotted the Barnes-Hut simulation energy spread along with the analytic prediction, which we have obtained by numerically integrating Eq. \ref{eq:ibs} using the beam size, beam length, and emittance arrays from the GPT simulation. We note that although the final energy spread is comparable between the two cases, the analytic growth rate seems to overestimate the IBS contribution in the drift prior to the waist and underestimates the contribution from the waist itself. 

\subsubsection{Validation of charge scaling}

\begin{figure}[h!]
    \centering
    \includegraphics{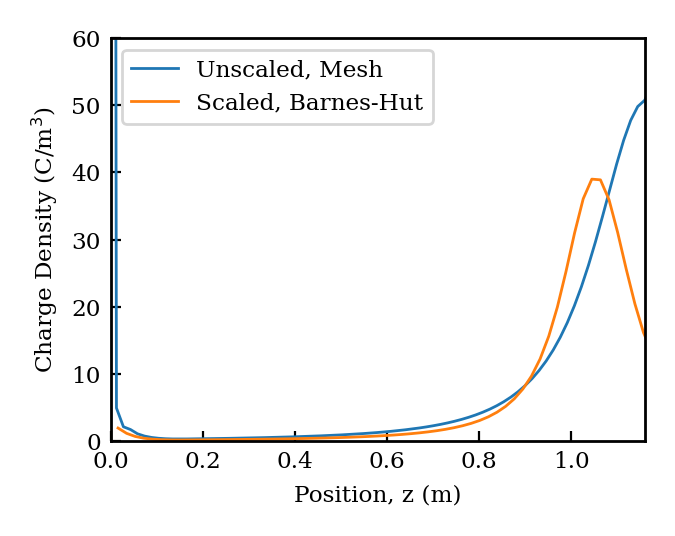}
    \includegraphics{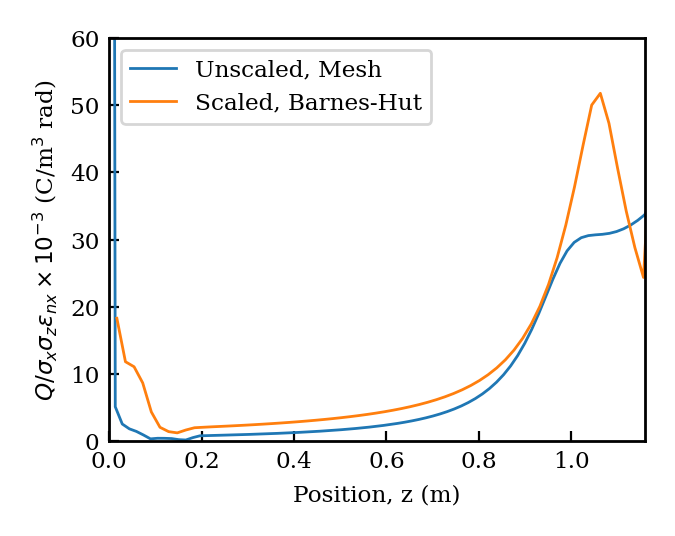}
    \caption{The two ratios of interest, the charge density (top) and $Q/\sigma_x\sigma_z\epsilon_{nx}$ (bottom), are plotted for the 100 pC mesh simulation and the 100 fC Barnes-Hut simulation.}
    \label{fig:ibs_validation}
\end{figure}

As previously discussed, short-range Coulomb effects intuitively scale roughly with the bunch charge density, hence our approach to making this problem tractable. More precisely we see from the theoretical predictions reproduced in Eq. \ref{eq:ibs} that the true scaling is not quite with the charge density but involves the beam angle through the emittance with the term $Q/\sigma_x\sigma_z\epsilon_{nx}$. For the sake of extrapolating the results of these scaled, 100 fC simulations to the original 100 pC gun, we should thus demonstrate that this ratio, as well as the beam density, is roughly preserved throughout the gun when the scaling is performed. We validate this numerical approach in Figure \ref{fig:ibs_validation}, where the charge density (top) and the ratio $Q/\sigma_x\sigma_z\epsilon_{nx}$ (bottom) are plotted for the 100 pC simulation and for the 100 fC Barnes-Hut simulation. As expected from the envelope equation and emittance compensation theory, there is good, though not perfect, agreement between the two simulations for both quantities of interest. Quantitatively, they differ from each other by no more than a factor of two at the transverse beam waist where short-range Coulomb interactions are strongest. Since the square of the energy spread has been shown to be proportional to these ratios, one may then interpret our 100 fC result as giving an estimate of the IBS contribution to the 100 pC gun energy spread within roughly 50\%. This additional energy spread, of roughly 600 eV magnitude, is added in quadrature with the energy spread from the 100 pC simulation.

\subsubsection{Emittance growth estimate from disorder-induced heating}

Finally, we will attempt to estimate the magnitude of the expected emittance growth from DIH in this scaled scenario. Associated with disorder induced heating is an increase in the effective transverse temperature of the electron beam \cite{Maxson_2013},
\begin{equation}
    k_BT_\mathrm{DIH} = \frac{0.45e^2}{4\pi\epsilon_0a},
\end{equation}
where $a=(4\pi n_b/3)^{-1/3}$ is the Wigner-Seitz radius associated with the electron number density $n_b$ at the cathode. This temperature increase is achieved on the timescale of a single plasma oscillation, similar to the emittance compensation dynamics themselves. This corresponds to an increase in the transverse beam emittance of magnitude
\begin{equation}
    \epsilon_\mathrm{DIH} = \sigma_x\sqrt{\frac{2k_BT_\mathrm{DIH}}{3mc^2}}.
\end{equation}
As what will be an overestimate of the DIH effect, let us use the parameters as they are at the cathode, although in reality the temperature increase is achieved over a single plasma period and thus samples something closer to the average, lower beam density. For the parameters at the cathode we find an effective DIH temperature increase of 4.2 meV, resulting in an emittance growth of 0.06 nm rad. This implies a negligible contribution to the beam dynamics, validating our emphasis on IBS effects. Indeed this magnitude is completely ignorable relative to the 10's of nm scale of the 100 pC emittance, however can become important as the charge is scaled down for potential UED and UEM applications.

\section{\label{sec:magbeam}Magnetized Operation for Asymmetric Emittance Beams}

In the following sections we describe an alternative operating point for the injector in which the photocathode is magnetized by an additional solenoid placed behind it. The beam generated at the cathode is imbued with canonical angular momentum: an emittance-like contribution to the transverse beam dynamics which is physically realized in linear correlations between the phase space variables $x-y'$ and $y-x'$. These correlations may be removed downstream of the injector via skew quadrupole magnets, but only after the radial emittance is properly compensated. The resulting beam can have asymmetric transverse emittances with $\epsilon_x/\epsilon_y$ on the scale of 100 while maintaining a four-dimensional emittance, $\epsilon_{\mathrm{4D}}=\sqrt{\epsilon_x\epsilon_y}$, on the scale of the ultra-high brightness FEL injector. Such a beam is attractive for several applications: at low charge this ``flat" beam is ideal for injection into a dielectric laser accelerator with a slab geometry. At higher charge, on the order of 1 nC, flat beams are demanded for a future linear collider in order to mitigate beamsstrahlung at the interaction point. In order to maximize the versatility of the design and physics we present here, we will identify an operating point with 100 pC charge which may be scaled up or down according to the requirements of the application. 

\subsection{Beam requirements for a future linear collider}

Linear collider designs require an asymmetric  beam at the final focus, where the ratio of horizontal to vertical beam size is nearly two orders of magnitude \cite{bambade2019international}, to mitigate the negative effects of the strong beam-beam interaction \cite{ChenTelnov}. This in turn implies an emittance ratio $\epsilon_x/\epsilon_y\simeq 100$, with the smaller emittance, $\epsilon_y$, at the the $10^{-8}$ m rad level. In order to the provide the needed collider luminosity, there should also be a significant charge in the beam.  Traditionally, such phase space requirements are met by damping rings, which naturally produce asymmetric beams. However, such rings are costly, and it has long been speculated that one might replace the electron damping ring with an asymmetric emittance-producing photoinjector. 

Production of asymmetric beams with a photoinjector is indeed enabled by strongly magnetizing the photocathode, and subsequently removing the beam's angular momentum with a skew-quadrupole array \cite{brinkmann2001low}. In the process, one may produce a large splitting of the transverse emittances. As we shall see below, the conservation laws at play imply that to reach linear collider-appropriate performance levels, one must press the state-of-the-art in beam brightness produced by the source. Thus the very high field photoinjector is of interest in this application. This approach requires an understanding of beam brightness limitations of a magnetized beam due the effect of beam angular momentum on
the emittance compensation process. Further, one must optimally perform the removal of angular momentum through a ``flat-beam transform". This second step is discussed first, as it sets the requirements for the first. 

\subsection{Flat beam transform of magnetized beams}

In principle there are several approaches to producing flat electron beams from photoinjectors. By utilizing a blade array geometry for the photocathode \cite{lawler2019electron} or by shaping the transverse laser profile \cite{rosenzweig1991flat}, for example, one can produce beams with large transverse emittance ratios directly. However, emittance compensation is difficult when the beam is not cylindrically symmetric, just one reason for which being that the Larmor rotation of the beam in subsequent solenoid magnets can cause growth of the projected emittances through unwanted coupling between transverse phase space planes. For an azimuthally symmetric beam this entails no deleterious effects, but for an asymmetric beam this rotation can spoil the beam emittance.

Partially on account of these limitations, in this design we utilize a robust approach based on introducing canonical angular momentum to the beam through immersion of  the cathode in an axial magnetic field using an external solenoid magnet \cite{PiotAsym}. This technique has the significant advantage that the beam maintains its cylindrical symmetry up until its emittance is fully compensated; only after space-charge effects are strongly diminished is the beam permitted spatial asymmetries in design. Insofar as tracking the transverse envelope dynamics is concerned, the angular momentum acts effectively like additional emittance, as noted above, added in quadrature to the thermal emittance of the cathode. Unlike the thermal emittance, however, the canonical angular momentum may be removed by skew quadrupole magnets placed downstream of the injector \cite{brinkmann2001low}, which introduce the asymmetries mentioned above, and serve to split the transverse emittances. Indeed, in the process of removing the cross-correlations in the transverse phase space, this transformation leaves the beam with a non-unity transverse emittance ratio which can in principle be quite large, often exceeding 100 \cite{PiotAsym}.

We begin the discussion of applying this scheme to the high field photoinjector by reviewing the basic principles of flat beam generation from a magnetized photocathode. An electron released with a non-vanishing axial magnetic field at the cathode $B_z(0)$ will have a conserved canonical angular momentum $L =eB_z(0)r^2/2$ such that a bunch of transverse size $\sigma_r$ is characterized by a canonical angular momentum value of $L=eB_z\sigma_r^2/2$. Such a beam is then characterized by split geometric \textit{eigen-emittances} $\epsilon_{\pm}=\sqrt{\epsilon_{\mathrm{4D}}^2+\mathcal{L}^2}\pm \mathcal{L}$ where $\epsilon_{\mathrm{4D}}=\sqrt{\epsilon_{x}\epsilon_{y}}$ is the geometric 4D emittance and $\mathcal{L}=L/2p_z$ is the geometric analogue of the conserved rms canonical angular momentum. In the limit that $\mathcal{L}\gg\epsilon_{\mathrm{4D}}$ the eigen-emittances are approximately $\epsilon_+=2\mathcal{L}$ and $\epsilon_-=\epsilon_{\mathrm{4D}}^2/2\mathcal{L}$. Thus the emittance ratio is $\epsilon_+/\epsilon_-=(2\mathcal{L}/\epsilon_{\mathrm{4D}})^2$. This implies a complicated tradeoff which goes into maximizing the transverse emittance ratio: a larger axial field on the cathode will increase the ratio, however it will also make compensation more difficult and potentially increase the effective $\epsilon_{\mathrm{4D}}$ in the final beam through poor compensation. Similarly, both the canonical angular momentum and the thermal emittance increase with the spot size on the cathode with the latter scaling linearly and the former quadratically.  

\subsection{Baseline injector performance}

The geometry of the injector for flat beam mode is similar to the ultra-high brightness mode with the addition of a secondary solenoid placed behind the cathode to provide an axial magnetic field on the cathode surface. This solenoid has a similar engineering philosophy as the original magnet described in Section \ref{sec:engineering}. We will distinguish between this and the existing solenoid by referring to the latter as the compensating solenoid. Additionally, the injector is followed by three skew quadrupole magnets which perform the flat beam transformation after the emittance is properly compensated.   

\subsubsection{Emittance compensation}

\begin{figure}[h!]
    \centering
    \includegraphics{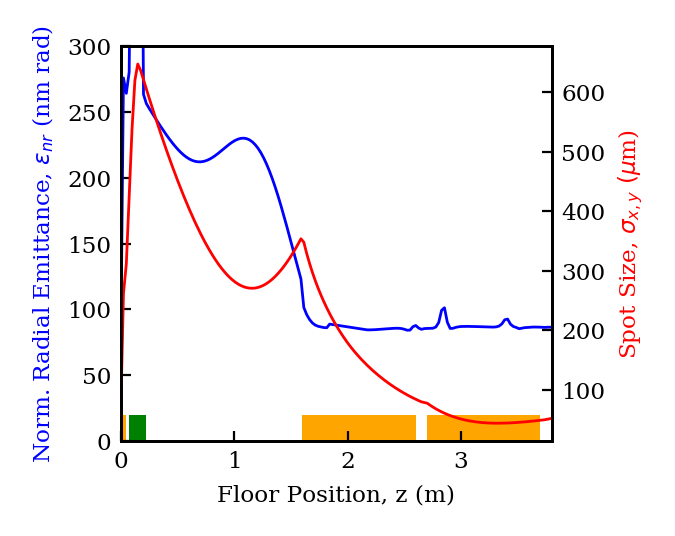}
    \caption{The emittance and beam size evolution of the magnetized operating point are shown up to the entrance to the skew quadrupole triplet.}
    \label{fig:magnetized_compensation}
\end{figure}

We will begin the discussion of the injector performance in the flat beam mode by summarizing the results of the emittance compensation. The relevant parameters for the operating point are given in Table \ref{tab:magnetized} and the compensation profile is plotted in Figure \ref{fig:magnetized_compensation}. The on-axis field profile is the same as in the ultra-high brightness case, see Figure \ref{fig:gunprofile}.

\begin{table}[h!]
    \centering
    \begin{tabular}{|c|c|c|}
    \hline
        Parameter & Unit & Value   \\
    \hline
        Charge & pC & 100\\
        Laser Spot Size (Pre-Cut) & $\mu$m & 151\\
        Laser Spot Size (Post-Cut)& $\mu$m & 76\\
        Injection Phase & $^{\circ}$ & 44\\
        Laser Length & ps & 5.8\\
        Peak Cathode Field & MV/m & 240\\
        \hline
        Bucking Solenoid Field & T & 0.58\\
        Compensation Solenoid Field & T & 0.48\\
        Compensation Solenoid FWHM & cm & 7.4\\
        Compensation Solenoid Center & cm & 12.5\\
        Booster Gradient & MV/m & 52\\
        Booster Entrance & m & 1.6\\
        Booster Phase & $^{\circ}$ & 90\\
        \hline
    \end{tabular}
    \caption{The injector parameters relevant to emittance compensation for magnetized beam operation are listed. }
    \label{tab:magnetized}
\end{table}

Since there are relatively few examples of magnetized beam compensation, we will take some time to discuss the features of this profile and compare it to our arguments in Section \ref{sec:envelope}. We note that upon exiting the compensating solenoid the beam is converging with an envelope angle $\sigma_x'=-620$ $\mu$rad. Neglecting space-charge defocusing, this corresponds to an angular momentum dominated waist beam size of $\sigma_{x,0}=-L/\gamma\sigma_x'=250$ $\mu$m, which is only slightly smaller than the waist realized in the simulation indicating that the angular momentum derived defocusing forces are comparable to those derived from space-charge during the drift prior to the linac. This is in notable contrast to the FEL operating point, for which an equivalent calculation using the normalized emittance instead of the angular momentum yields a 7 $\mu$m beam size. This is very small relative to the 63 $\mu$m spot realized in the FEL injector simulation, which is indicative of the highly dominant space-charge defocusing effects in that scenario. Note that the implications of this are interesting: the beam sizes associated with space-charge are naturally much smaller than those associated with angular momentum when the angular momentum is this large even when the two forces are strictly equal in magnitude. This justifies a perturbative approach to solving for the envelope dynamics even when angular momentum is not completely dominant.  

It is worth noting that the booster linac entrance is not coincident with the location of the first emittance maximum as it usually is and was in the FEL operating mode. This was foreshadowed in Section \ref{sec:envelope} -- the large angular momentum contribution makes it such that the beam becomes angular momentum dominated at a much lower energy. Subsequently, for similar accelerating gradient one must place the booster linac closer to the final emittance minimum. If one tries to inject at the waist the required accelerating gradient will be too low and the beam will not properly focus. It is worth noting that in the case presented here, the emittance is frozen almost exactly at the value when the beam size drops below the constant beam size solution associated with the angular-momentum-dominated dynamics $\sigma=(8/\eta)^{1/4}(L/\gamma')^{1/2}\approx 210$ $\mu$m.

We have also included in the design a second 1 m linac structure located 10 cm downstream of the initial booster linac. The emittance compensation process is complete before the beam enters this additional linac as indicated by the flat emittance profile between the two linacs. The second linac is placed before the skew quadrupole configuration for two purposes. First, the beam energy at the exit of the booster is relatively low at just 55 MeV. This relatively low energy beam is susceptible to space-charge effects during the very sensitive skew quadrupole transformation; we thus use a second structure operated at the maximally allowed accelerating gradient to increase the energy to 210 MeV. Furthermore, the second linac provides additional focusing to the beam to minimize deleterious nonlinear effects associated with the second-order quadrupole transfer matrix components.

The final normalized radial emittance produced before the skew quadrupoles is quite good, reaching a minimum value of 85 nm rad as shown in Figure \ref{fig:magnetized_compensation}. Based on this we may estimate an optimal transverse emittance ratio of nearly 500 with $\gamma\epsilon_+=1.9$ $\mu$m rad and $\gamma\epsilon_-=3.8$ nm rad. 

\subsubsection{Skew quadrupole transformation}

The second linac is followed immediately by a skew quadrupole triplet (SQT). We model the magnets as 8 cm long with their consecutive edges placed 1 m apart. The first magnet edge is located 4 m downstream of the cathode. The gradients are -14.53 T/m, 9.27 T/m, and -5.36 T/m, respectively. These optimal values were obtained initially using \texttt{elegant} \cite{borland2000elegant} for fast optimization, then simulated in GPT to account for 3D transverse space-charge effects.

\begin{figure}[h!]
    \centering
    \includegraphics{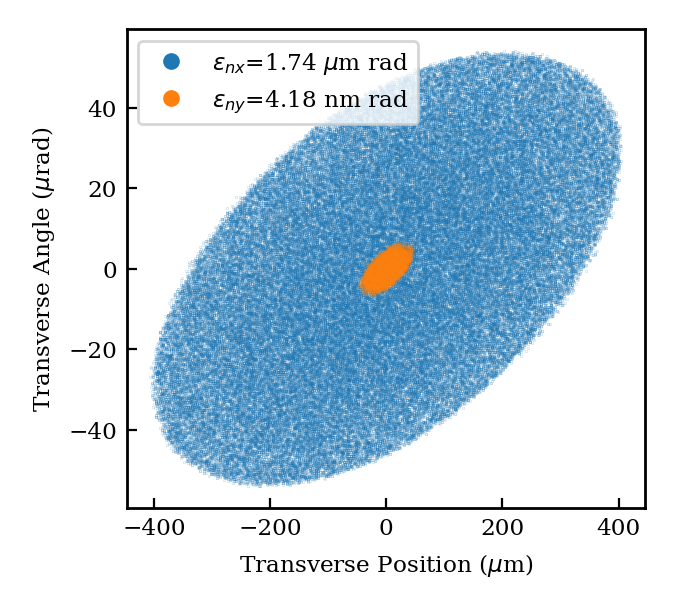}
    \caption{The transverse phase spaces are plotted on top of each other after the emittances have been split by the skew quadrupole triplet.}
    \label{fig:split_phasespaces}
\end{figure}

\begin{figure}[h!]
    \centering
    \includegraphics{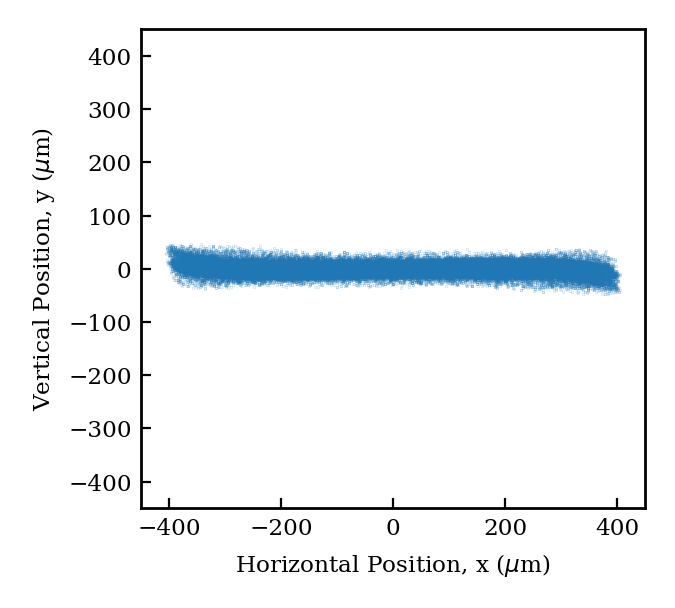}
    \caption{The x-y projection of the beam is shown after the skew quadrupole triplet.}
    \label{fig:split_projection}
\end{figure}

The transverse phase spaces after the SQT are plotted on top of each other in Figure \ref{fig:split_phasespaces}. As indicated in the figure, the SQT leaves the normalized transverse emittances split between 4.2 nm rad in x and 1.74 $\mu$m rad in y. This amounts to a transverse emittance ratio of 414, which is in excess of the requirements of the linear collider. This falls short of the factor of 500 predicted by our simple theoretical estimates. We have determined through comparison of \texttt{elegant} and GPT results that roughly 0.5 nm rad of the final smaller emittance is directly derived from a combination of space-charge effects during the skew quadrupole transformation, chromatic effects owing to the nonlinear longitudinal phase space, and nonlinearities in the quadrupole magnetic fields, none of which are accounted for in the theory employed for the estimate. If this approximately 0.5 nm rad contribution is ignored we retrieve roughly the theoretical prediction of $3.8$ nm rad for the smaller emittance. Further, we show in Figure \ref{fig:split_projection} the x-y projection of the beam. 

\subsection{Charge scaling}

As we saw in the previous sections, the presented injector design is highly versatile as one can move between an ultra-high brightness FEL operating point and a magnetized asymmetric emittance operating point with relatively minor adjustments to the gun operating conditions. This versatility has been demonstrated thus far for a 100 pC beam. 100 pC is ideal for FEL applications, but it may be too low to be practical in a linear collider scenario and too high for DLA. For a linear collider, at least in common present-day designs, one would like to obtain an operating point with a similar emittance ratio but with 1 nC bunch charge. Here we will discuss the relevant fundamental photoinjector scaling laws that govern such an increase to the bunch charge. We will conclude with a discussion of the corresponding challenges for scaling down to the single pC scale for DLA.

The standard approach for scaling the charge of a photoinjector design is to correspondingly scale the beam dimensions at the cathode such that the beam density is unaffected, thereby preserving the plasma frequency which sets the length scale for emittance compensation. In this way we may say that the transverse beam size and the beam length at injection will scale with the cube root of the beam charge. Recall also that the canonical angular momentum scales as the square of the beam size on the cathode, $\mathcal{L}\propto \sigma_x^2$ whereas the thermal emittance scales linearly with the same quantity. Under the assumption that the final emittance is nearly equal to the thermal emittance, an assumption we will study in more detail further below, we may conclude that $\epsilon_{\mathrm{4D}}\propto \sigma_x$. As a result, the emittance ratio will scale as $\epsilon_+/\epsilon_-=(2\mathcal{L}/\epsilon_{\mathrm{4D}})^2\propto\sigma_x^2\simeq Q^{2/3}$. Thus, one might expect that increasing the bunch charge by an order of magnitude would in fact improve the final emittance ratio by roughly $10^{2/3}$. Meanwhile, the smaller emittance $\epsilon_-=\epsilon_{\mathrm{4D}}^2/2\mathcal{L}$ is insensitive to the charge scaling in this limit, so the sub-10 nm smaller emittance would also be preserved in accordance with the requirements for a collider. 

This analysis holds only under the original assumption that the final emittance scales with spot size in the same way as the thermal emittance, which is not necessarily the case. In particular, lengthening the bunch at the cathode may enhance the emittance contribution from the rf wave, which is caused by different portions of the beam sampling different phases of the accelerating wave at the exit iris thereby picking up a time-dependent angular kick. This effect is represented by a contribution to the emittance of
\begin{equation}
    \epsilon_{x}^\mathrm{rf} = \frac{eE_0}{2mc^2}\sigma_x^2k_\mathrm{rf}^2\sigma_z^2,
\end{equation}
which is added in quadrature to the other contributions to the total emittance. For the operating conditions in the magnetized photocathode scenario scaled according to the prescription described above, this contribution comes to be about 750 nm rad, much larger than the sub-100 nm rad thermal emittance. Therefore, one expects that in this case the four-dimensional emittance will be dominated by the rf contribution, leading to the modified scaling $\epsilon_{\mathrm{4D}}\propto \sigma_x^2\sigma_z^2\propto Q^{4/3}$. Correspondingly, the emittance ratio scales as $Q^{-4/3}$ and the smaller emittance as $Q^{2/3}$. The new smaller emittance will still be of order 10 nm rad, however the ratio will be reduced to be of order 10 itself, which is insufficient for the collider. The most obvious way to resolve this issue is by lengthening the rf wavelength, perhaps by moving instead to S-band (as studied in \cite{rosenzweig2019next}) where the rf frequency is a factor of 2 smaller and thus the rf emittance is reduced by a factor of four assuming all other variables are held constant. Indeed, such a change reduces the rf emittance to the 200 nm rad level. This prediction corresponds to experience; when one is attempting to produce nC-class high brightness beams, one should operate at an rf wavelength of at least 10 cm.   

In the opposite case, one might naturally imagine scaling down the bunch charge to appeal to DLA applications. In this case rf effects only become weaker, so we may return to the scaling arguments made previously finding $\epsilon_+/\epsilon_-\propto Q^{2/3}$. Thus the emittance ratio is reduced by a reduction in the charge, however for the DLA the quantity of higher performance is the smaller emittance $\epsilon_-=\epsilon_{\mathrm{4D}}^2/2\mathcal{L}$ which is independent of the beam charge in this limit. The smaller emittance is the more important quantity here because it determines whether the beam can be properly matched into the DLA section: the size of the beam in the larger plane is largely irrelevant. We note that the minimal emittance observed in the current injector design, 4 nm rad, is consistent with the constraints noted in \cite{cropp2019maximizing} required for DLA experiments at the UCLA Pegasus Lab. It is worth noting also, however, that the reduction in the emittance ratio makes it such that this scaling is only worthwhile when the photocathode itself cannot be operated with sufficiently low thermal emittance. 

\section{Conclusions}

We have presented here a versatile, cryogenically-cooled photoinjector which promises beams of unprecedented six-dimensional brightness. The fundamental innovation allowing this step towards ever brighter beams is the use of normal conducting rf cavities at cryogenic temperatures to launch the beam from the cathode at extremely high fields. This work for the first time has considered a feasible engineering design for all injector components, in the process revealing as of yet largely unconsidered physics concepts related to the presence of non-resonant spatial harmonics in the rf field. Indeed we have found that the inclusion of these harmonics, demanded by the structure's power efficiency, serves to enhance the emittance compensation by providing stronger focusing on the beam both as it leaves the cathode and as it is accelerated towards an emittance dominated state downstream, yielding an unprecedented 45 nm rad emittance at 100 pC and 20 A level current. With the cryogenic nature of the gun comes additional engineering difficulties with the solenoid required to facilitate emittance compensation, which we have addressed with a design for a cryo-solenoid which can rest in the same cryostat as the gun.

In addition to exploring critical new aspects of the concept of a high-field cryogenic injector, we have presented a unique study of the effects of microscopic collective beam dynamics on the performance of a photoinjector. Our approach, which takes advantage of the fundamental principles driving these short-range Coulomb interactions, has allowed us to estimate the scale of the energy spread incurred by a beam due to intra-beam scattering in an injector -- an estimate which is not just novel but also critical for the feasibility of an ultra-compact x-ray free-electron laser using cryo-rf accelerators. 

We have further shown that the same fundamental injector design may be modified by the inclusion of a photocathode-magnetizing solenoid behind the gun to realize beams with unprecedentedly small, asymmetric transverse emittances and a comparably small four-dimensional emittance. This entails a uniquely detailed study of the process of emittance compensation in the presence of canonical angular momentum. The beam demonstrated in these simulations -- with 100 pC charge, an emittance ratio of 400, and a smaller emittance of 4 nm rad -- sheds light on the efficiency of emittance compensation in this regime and additionally provides a first step for developing a scaled beam at a larger charge suitable, for injecting into a linear collider. 

The experimental realization of this gun is in progress at the UCLA SAMURAI laboratory, for testing both FEL and linear collider applications. Before commissioning this gun, early tests will seek to demonstrate the efficiency gains induced by using a high-field, strongly focusing distributed coupling linac as the booster for an existing injector - the UCLA hybrid gun \cite{FUKASAWA2012}. Simultaneously, UCLA is developing a 0.5 cell C-band test cavity for dedicated study of the properties of beams emitted from cryogenically-cooled photocathode surfaces. This naturally includes the possibility of studying cryo-emission in depth, in addition to the behavior of cryogenically cooled cathodes exposed to high intensity UV lasers. Developing a proper understanding of the details of emission in this physical scenario is critical for the small spot size operation we envision. UCLA has acquired a 5 MW C-band klystron, which is being commissioned and is capable of feeding this half-cell structure. With this variety of beam sources soon to be available at the SAMURAI lab, early experimental work will also focus on demonstrating the utility of these linac structures and gun designs in permitting robust FEL lasing. The first iterations of this will entail lasing at optical and EUV wavelengths with a several hundred MeV beam. Such an initial step allows for early investment into the other enabling technologies of the UC-XFEL, such as meso-scale cryo-undulators and IFEL modulation-based compression schemes. These first steps will enable the realization of the full UC-XFEL, producing soft x-ray photons with a 1 GeV electron beam in just 40 m.   

\section*{Acknowledgements}

This work was performed with support of the National Science Foundation through the Center for Bright Beams, Grant No. PHY-1549132. Support was also obtained from the US Dept. of Energy, Division of High Energy Physics, under contracts no. DE-SC0009914 and DE-SC0020409.

\appendix

\section{\label{sec:fieldprof}Optimal $\pi$-mode field profile for gradient enhancement}

The acceleration of ultra-relativistic particles in a standing wave cavity has the interesting property that the rate of energy gain by the particles is sensitive only to the fundamental spatial harmonic, while the walls of the cavity itself are sensitive primarily to the net field strength. As indicated in previous sections, this implies that the accelerating gradient observed by an ultra-relativistic particle may be higher than the peak longitudinal electric field in the cavity if the field contains higher-order spatial harmonic content. This in turn suggests that structures which support high harmonic content will naturally be more efficient at supporting high gradients, a claim which seems to have been validated by the numerical optimizations of \cite{Tantawi2020} which yielded our particular field profile. In this section we will demonstrate that the theoretical limit of this effect is achieved by a perfect square wave field profile, for which the accelerating gradient exceeds the expected value by the factor $4/\pi$.

To start we should formally state the problem to be solved. This is to determine for what valid $\pi$-mode field profile is the maximum value of the field a minimum relative to the strength of the fundamental accelerating wave. We propose that the answer is a square wave defined according to 
\begin{equation}
    E_0(x) = \begin{cases}
    2/\pi & -\pi/2<x<\pi/2\\
    -2/\pi & \pi/2<x<3\pi/2
    \end{cases}.
\end{equation}
Of course, a $\pi$-mode field profile is any field profile which may be expressed in the form of an odd-frequency cosine series, 
\begin{equation}
    E(x) = \sum_{n=0}^\infty a_{2n+1}\cos((2n+1)x),
\end{equation}
in which we force the normalization $a_1=1$. We will prove that the square wave $E_0(x)$ is the answer to the posed problem by contradiction, by supposing that in fact another field has a smaller maximum than the square wave when both have $a_1=1$. We will define this field according to its difference from the square wave,
\begin{equation}
    E(x) = E_0(x) + E_1(x),
\end{equation}
where by necessity $E_1(x)$ is itself expressible as an odd-frequency cosine series,
\begin{equation}
    E_1(x) = \sum_{n=0}^\infty b_{2n+1}\cos((2n+1)x).
\end{equation}
Since the square wave achieves its maximum value at every location in the range $(-\pi/2,\pi/2)$, $E(x)$ can only have a lower maximum if $E_1(x)$ is negative everywhere in this range. The reason is simply that if at any point $E_1(x)$ is positive, then $E(x)$ will at that point exceed the constant maximum value of the square wave. Simultaneously however, $E(x)$ must maintain a fundamental wave strength of 1. Since $E_0(x)$ itself already has fundamental coefficient equal to 1, we must have that the corresponding coefficient of $E_1(x)$ vanishes. In other words, 
\begin{equation}
    b_1 = \frac{1}{\pi}\int_{-\pi/2}^{3\pi/2}E_1(x)\cos(x)dx = 0.
\end{equation}
Since both $E_1(x)$, on account of its form, and $\cos(x)$ have the property that $f(x+\pi)=-f(x)$, this is equivalently written as 
\begin{equation}
    b_1 = \frac{2}{\pi}\int_{-\pi/2}^{\pi/2}E_1(x)\cos(x)dx.
\end{equation}
Of course, if $E_1(x)$ is negative across the entire integration range, then this cannot vanish. Thus $E_1(x)$ must become positive at some point in this range, thereby granting $E(x)$ a maximum which is larger than that of the square wave. 

With this fact established we may consider its ramifications for an ideal - under the present consideration - $\pi$-mode field profile. The Floquet form of a square wave accelerating field is 
\begin{equation}
    E(z) = 2E_0\sum_{n=0}^\infty \frac{(-1)^n}{2n+1}\cos((2n+1)k_\mathrm{rf}z).
\end{equation}
It follows that the maximum of the field profile is related to the amplitude of the fundamental wave by 
\begin{equation}
    E_\mathrm{max} = 2E_0\sum_{n=0}^\infty \frac{(-1)^n}{2n+1} = \frac{\pi}{2}E_0.
\end{equation}
The expected accelerating gradient from a purely sinusoidal field profile is half of the peak field, so we conclude that the accelerating gradient $E_0$ for an ultra-relativistic particle is $E_0=(4/\pi)(E_\mathrm{max}/2)$, implying an enhancement over the equivalent purely sinusoidal accelerating gradient of $4/\pi\approx 1.27$. Stated more explicitly, the square wave field profile achieves a 27\% higher accelerating gradient for the same peak field strength relative to a pure first harmonic wave. In particular, this implies a 27\% higher allowed gradient before rf breakdown becomes problematic.

\section{\label{sec:invenvsolutions}Absence of invariant envelope type solutions in magnetized photoinjectors}

The invariant envelope is attractive as a particular solution to the envelope equation in a booster linac largely due to the invariance of the associated phase space angle with respect to local current. This guarantees that as the beam is accelerated, first-order correlations in the phase space are removed by the time the beam becomes emittance-dominated. Ideally one would like to find an analogous solution in the presence of angular momentum, however we will show here that no such solution exists. This is not to say that there are not solutions which facilitate compensation, just that there is no solution with the property that $m(z)=\sigma'(z,I)/\sigma(z,I)$ is independent of the current, and therefore it is more difficult to guarantee efficient compensation.

Proving this statement begins by replacing $\sigma'(z)$ in the envelope equation by the trace space angle $m(z)$ defined above. This yields
\begin{equation}
\begin{split}
    m'(z) + m(z)^2+\frac{\gamma'}{\gamma}m(z)+\frac{\eta}{8}\left(\frac{\gamma'}{\gamma}\right)^2 =\\
    \frac{L^2}{\gamma^2\sigma(z,\zeta)^4}+\frac{I}{2I_0\gamma^3\sigma(z,\zeta)^2}
\end{split}
\end{equation}
If we ask that the trace space angle not vary with current, then the left-hand side is independent of $\zeta$. As a result we must also have the right-hand side independent of the current. Let us define 
\begin{equation}
    g(z) = \frac{L^2}{\gamma^2\sigma(z,\zeta)^4}+\frac{I}{2I_0\gamma^3\sigma(z,\zeta)^2}
\end{equation}
where the function $g(z)$ is explicitly independent of the current. This expression can be inverted to write the beam size in terms of this function,
\begin{equation}
    \sigma(z,\zeta)=\sqrt{\frac{I}{4I_0g(z)\gamma^3}\left[1+\sqrt{1+\frac{8I_0g(z)L^2\gamma^4}{I}}\right]}
\end{equation}
Under our original assumptions, the trace space angle corresponding to this envelope solution should satisfy $dm/dI=0$. Enforcing this condition returns a condition on $g(z)$, 
\begin{equation}
    \frac{dg}{dz} = -\frac{4g\gamma'}{\gamma}
\end{equation}
This of course can be solved simply
\begin{equation}
    g(z) = g(0)\left(\frac{\gamma_i}{\gamma}\right)^4
\end{equation}
If one plugs this back into the envelope, and subsequently into the envelope equation, there is no solution. Thus, the envelope equation with angular momentum does not permit solutions with constant phase space angles throughout the booster linac. 


\bibliography{apssamp}

\end{document}